\documentclass[twocolumn,english,aps,prb,showpacs]{revtex4}
\usepackage[T1]{fontenc}
\usepackage[latin9]{inputenc}
\usepackage{color}
\usepackage{textcomp}
\usepackage{mathrsfs}
\usepackage{amsmath}
\usepackage{amssymb}
\usepackage{graphicx}
\usepackage{esint}

\makeatletter

\providecommand{\tabularnewline}{\\}

\@ifundefined{textcolor}{}
{%
 \definecolor{BLACK}{gray}{0}
 \definecolor{WHITE}{gray}{1}
 \definecolor{RED}{rgb}{1,0,0}
 \definecolor{GREEN}{rgb}{0,1,0}
 \definecolor{BLUE}{rgb}{0,0,1}
 \definecolor{CYAN}{cmyk}{1,0,0,0}
 \definecolor{MAGENTA}{cmyk}{0,1,0,0}
 \definecolor{YELLOW}{cmyk}{0,0,1,0}
 }

\@ifundefined{definecolor}
 {\usepackage{color}}{}
\usepackage{babel}

\usepackage{babel}

\makeatother

\usepackage{babel}
\begin{document}

\title{A 3D topological insulator quantum dot for optically controlled quantum memory and quantum computing}

\author{Hari P. Paudel}

\author{Michael N. Leuenberger }

\email{michael.leuenberger@ucf.edu}

\affiliation{NanoScience Technology Center and Department of Physics, University of Central Florida, Orlando, Florida 32826, United States}
\begin{abstract}
We present the model of a quantum dot (QD) consisting of a spherical
core-bulk heterostructure made of 3D topological insulator (TI) materials,
such as PbTe/Pb$_{0.31}$Sn$_{0.69}$Te, with bound massless and helical
Weyl states existing at the interface and being confined in all three
dimensions. The number of bound states can be controlled by tuning
the size of the QD and the magnitude of the core and bulk energy gaps,
which determine the confining potential. We demonstrate that such
bound Weyl states can be realized for QD sizes of few nanometers.
We identify the spin locking and the Kramers pairs, both hallmarks
of 3D TIs. In contrast to topologically trivial semiconductor QDs,
the confined massless Weyl states in 3D TI QDs are localized at the
interface of the QD and exhibit a mirror symmetry in the energy spectrum.
We find strict optical selection rules satisfied by both interband
and intraband transitions that depend on the polarization of electron-hole
pairs and therefore give rise to the Faraday effect due to Pauli exclusion
principle. We show that the semi-classical Faraday effect can be used
to read out spin quantum memory. When a 3D TI QD is embedded inside
a cavity, the single-photon Faraday rotation provides the possibility
to implement optically mediated quantum teleportation and quantum
information processing with 3D TI QDs, where the qubit is defined
by either an electron-hole pair, a single electron spin, or a single
hole spin in a 3D TI QD. Remarkably, the combination of inter- and
intraband transition gives rise to a large dipole moment of up to
$450$ Debye. Therefore, the strong-coupling regime can be reached
for a cavity quality factor of $Q\approx10^{4}$ in the infrared wavelength
regime of around $10\:\mu$m.

\end{abstract}


\maketitle

\section{Introduction}

3D TIs are narrow-bandgap materials with topologically protected gapless
surface/interface states that are characterized by the linear spectrum
of massless Weyl fermions.\cite{key-1,key-2} In such materials, the
spins of the Kramers pairs are locked at a right angle to their momenta
on the Fermi surface due to spin-orbit coupling,\cite{key-7,key-16,key-17,key-18,key-15,key-19,Yusheng}
which can be used for spin current generation.\cite{key-6,key-13,key-14}
The surface states are protected by time reversal symmetry, leading
to suppression of backscattering from edges and nonmagnetic impurities.\cite{key-1,key-2,key-3,key-4,key-15}
Such states are of great importance in low-power opto-spintronics.\cite{key-5,key-6}
Decoherence can be circumvented by highly polarized spin states with
helical spin texture,\cite{key-7,key-8,key-9,key-10} leading to a
phase coherence length of several hundred nanometers in nanostructures.\cite{key-11,key-12}

In 3D TI nanostructures the special properties of topologically protected
surface states of TIs are amplified because of the large surface-to-volume
ratio. In addition, the chemical potential can be electrically tuned
using a gate voltage. For example, the coherent propagation of the
Weyl electrons around the perimeter of a nanoribbon provides excellent
evidence of the topological nature of the surface states in TI nanostructures.\cite{key-12}
Experiments on both the physical and chemical synthesis of TI nanostructures
have been done recently to understand their transport properties at
the nanoscale.\cite{key-27,key-28,key-29} Recently, in a TI QD with
tunable barriers based on ultrathin Bi$_{2}$Se$_{3}$ films, Coulomb
blockade with around 5 meV charging energy was observed.\cite{key-30}

So far, a theoretical study of electronic properties of 2D helical
states occurring at the nanoscale of 3D TIs, such as in QDs, is still
lacking. In this article, we present for the first time the study
of bound Weyl states that are confined at the interface of a spherical
core-bulk heterostructure QD made of 3D TI materials such as Pb$_{1-x}$Sn$_{x}$Te.
We show that at the interface massless Weyl fermions are confined
in all three dimensions. The directions of spin and momentum are tangent
to the surface of the QD. 
Remarkably, their inherent spin-momentum locking property exists even
in a QD. Because of the linear dispersion there is a mirror symmetry
in the energy spectrum between positive and negative energy states,
in contrast to topologically trivial semiconductors. We demonstrate
that this symmetry in energy spectrum is preserved for the QD spectrum.

\begin{figure}
\includegraphics[width=8.5cm]{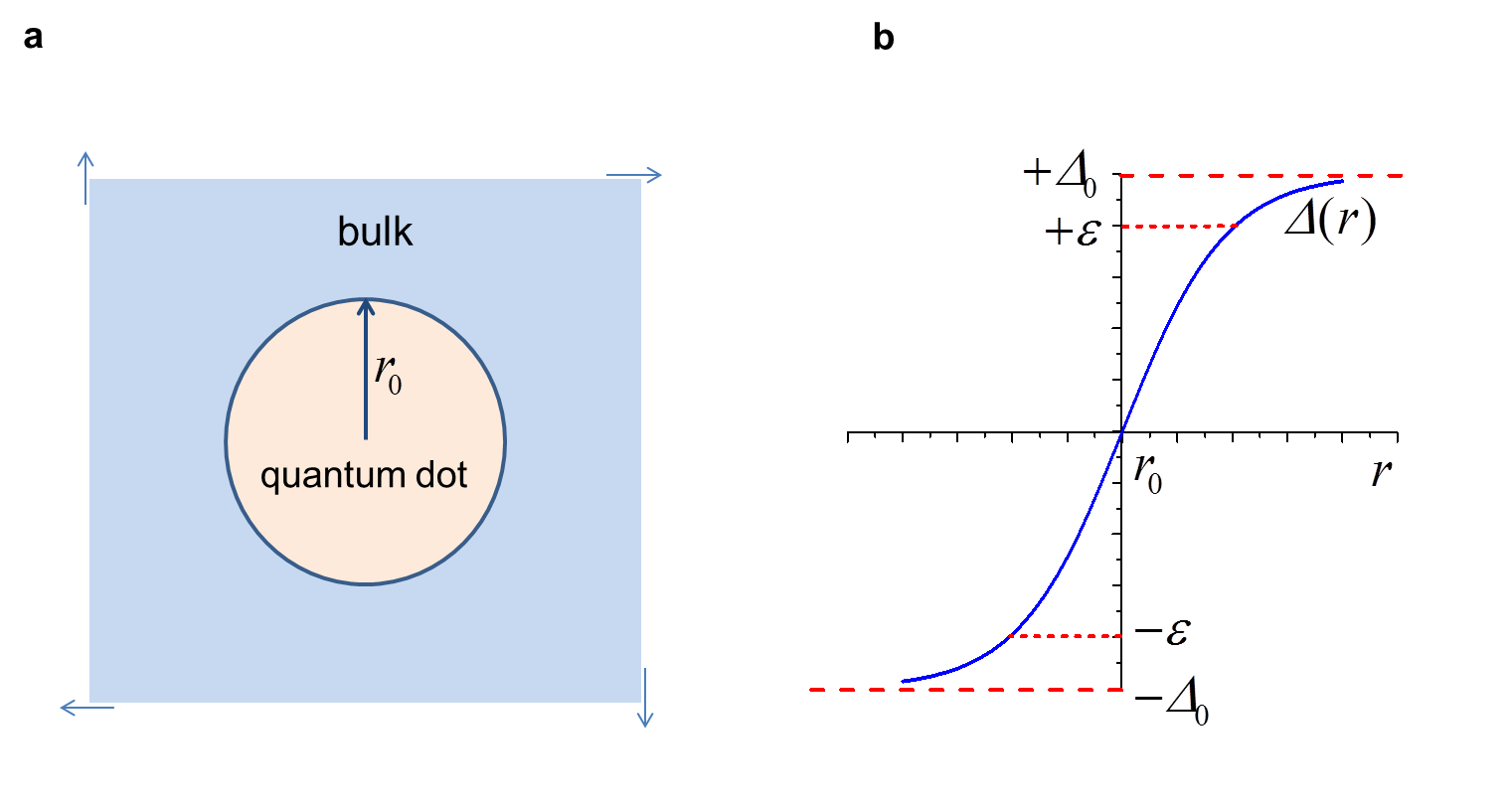}\caption{A heterostructure spherical core-bulk 3D TI QD with a single interface.
\textbf{a}. The arrows indicate the infinite size of the host. The
core and bulk host can be chosen as PbTe and Pb$_{0.31}$Sn$_{0.69}$Te
or vice versa. \textbf{b}. The potential $\Delta(r)$ binds Weyl fermions
at the interface. The energy of the bound interface states depends
on the size of the QD and the strength of the potential. \textcolor{black}{As
an example, two bound states at the interface are shown with energies
}$+\varepsilon$ \textcolor{black}{and }$-\varepsilon$\textcolor{black}{{}
(short dashed lines) for a QD of size $r_{0}=$2 nm. \label{QuantumDot} }}
\end{figure}

Several methods have been proposed to implement optically controlled quantum memory and optically mediated quantum computing with topologically trivial QDs. Quantum memories have been recently reviewed in Ref.~\onlinecite{Simon:2010}. A recent review on optically controlled quantum computing with electron spins can be found in Ref.~\onlinecite{Liu:2010}.
Optically controlled single-electron spin memory has been experimentally demonstrated using GaAs QDs\cite{Kroutvar:2004} and
InGaAs QDs.\cite{Ebbens:2005} Exciton memory has been implemented experimentally in a semiconductor nanopost.\cite{Krenner:2008}
For the purpose of using a hole spin as quantum memory or qubit, high coherence of hole spins in InGaAs QDs has been experimentally shown.\cite{Brunner:2009}
Ref.~\onlinecite{Atature:2007} demonstrates experimentally that a single spin can be read out using Faraday rotation.
Schemes for optically controlled two-qubit interaction have been proposed that are based on the exchange of virtual photons inside a cavity,\cite{Imamoglu:1999}
the optical RKKY interaction,\cite{Piermarocchi:2002}
dipole-dipole interaction,\cite{Calarco:2003}
Substantial experimental progress has been made to implement optically controlled electron spin state preparation,\cite{Atature:2006} hole spin state preparation,\cite{Gerardot:2008} single-spin readout,\cite{Vamivakas:2010} dephasing protection,\cite{Greilich:2006} two-qubit gate,\cite{Stinaff:2006,Robledo:2008} two QD-spin entanglement,\cite{Kim:2011} and spin-photon entanglement.\cite{Vamivakas:2009}

In Refs. \onlinecite{Leuenberger:2005,Leuenberger:2006} we developed the method of the Faraday rotation
of a single photon due to the Pauli exclusion principle occurring on a topologically trivial QD. 
Our proposed method can be used for entangling remote excitons, electron spins, and hole spins.
We showed that this entanglement can be used for the implementation of optically mediated quantum
teleportation and quantum computing.
Our ideas and methods have been plagiarized in Ref. \onlinecite{Hu:2008}.

Here we show that classical and single-photon Faraday rotation due to the Pauli exclusion principle in a 3D TI QD occur
due to strict optical selection rules satisfied by both interband
and intraband transitions that depend on the polarization of electron-hole pairs.
Based on this finding we propose that 3D TI QDs can be used as quantum memory and for the implementation of optically mediated quantum teleportation and quantum computing.
First, we propose that a single e-h pair in a 3D TI QD can be used as a quantum memory.
The information is stored in form of the polarization state of the e-h pair.
In order to be able to read out this information multiple times, we develop the method of Faraday rotation of a classical electromagnetic field due to Pauli exclusion principle in a 3D TI QD.
Second, we propose that the polarization of a single e-h pair, a single electron spin, or a single hole spin can be used as a qubit
in a 3D TI QD for the implementation of optically mediated quantum teleportation and quantum computing.
We develop the method of single-photon Faraday rotation in a 3D TI QD, which creates the entanglement between a single photon
and a qubit on the 3D TI QD. This entanglement is the resource for the implementation of quantum teleportation and quantum computing.

In wide bandgap semiconductor QDs optical inter- and intraband transitions are energetically separated because the bandgap is typically much larger than the QD level spacing.\cite{Singh} In contrast to that, we show that in 3D TI QDs inter- and
intraband transitions combine because of the vanishing bandgap at band crossing. The resulting large dipole moment of up to
$450$ Debye provides the possibility to reach the strong-coupling regime
for a cavity quality factor of $Q\approx10^{4}$ in the infrared wavelength
regime of around $10\:\mu$m.

The paper is organized as follows. In Sec.~\ref{sec:Model-Based-on}
we present the analytical derivation of the Weyl solution of the radial
Dirac equation using Greens function technique at the bulk-quantum
dot interface. The resulting eigenvalues and eigenfunctions are analyzed
in the Sec.$\:$\ref{sec:Bound-States-of}. The Sec.$\:$\ref{sec:Optical-Excitations}
is devoted to the evaluation of the optical transition matrix elements
and the discussion on them. We also discuss on the potential applications
of our results. In Sec. \ref{Faraday-Effect for QD} we explain the
Faraday rotation effect achieve in the 3D TI QD. The application of
the 3D TI QD as a quantum memory is explained in the Sec. \ref{sec:Quantum-Spin-Memory}.
where we also explain the Stark energy shift that can be used to achieve
clean selection rules for the excitation of a single electron-hole (e-h) pair. The Sec. \ref{Single-Photon-Faraday-Effect}
and \ref{Quantum-Teleportation} are devoted to the detailed description
of the single-photon Faraday effect, where we show that a single
e-h pair, a single electron, or a single hole can be used as a qubit
to implement optically mediated quantum teleportation and quantum
computing with 3D TI QDs. In this section, different possible level
configurations of the Weyl states are shown to achieve the Faraday
rotation effect.

\section{\label{sec:Model-Based-on}Model Based on Dirac Equation}

In Fig.~\ref{QuantumDot} we show the model of our spherically symmetric
3D TI QD of a core-bulk structure with a single interface at radius
$r=r_{0}$. This core-bulk structure consists, for example, of an
inner core of PbTe and an outer bulk of Pb$_{0.31}$Sn$_{0.69}$Te
with bandgaps of 0.187 and -0.187 eV, respectively, or vice versa,
so that Weyl fermions are generated at the interface. Here we used
the bandgap formula provided in Ref. 24 for determining $x$. Note
that the band crossing happens in Pb$_{1-x}$Sn$_{x}$Te at $x=0.35$
at{} 4 K. The Weyl fermions are subjected to the spherically symmetric
potential $\Delta(r)$ (Fig.~\ref{QuantumDot} (b)).

To understand the properties of a 3D TI QD, we start with the Dirac
Hamiltonian within the $\mathbf{k}\cdot\mathbf{p}$ approximation.\cite{Nimtz:1983}
Neglecting the far band terms, we have 
\begin{equation}
H=v_{\Vert}\alpha_{z}\hat{p}_{z}+v_{\bot}\boldsymbol{\alpha}_{\bot}\cdot\mathbf{\hat{p}}+\beta\Delta\label{eq:1}
\end{equation}
 where $\boldsymbol{\alpha}=\left(\begin{array}{cc}
0 & \boldsymbol{\sigma}\\
\boldsymbol{\sigma} & 0
\end{array}\right)$ are the Dirac $\boldsymbol{\alpha}$- matrices, $\mathbf{\boldsymbol{\sigma}}$
are the Pauli matrices, $\beta=\left(\begin{array}{cc}
1 & 0\\
0 & -1
\end{array}\right)$ is the Dirac $\beta$-matrix, and $\mathbf{\mathbf{\hat{p}}}$ is
the momentum operator. The component of the Fermi velocities $v_{\bot}$and
$v_{\Vert}$ in angular and radial direction are determined by the
$v_{\bot}=P_{\bot}/m_{0}$ and $v_{\Vert}=P_{\Vert}/m_{0}$ respectively,
where $P_{\bot}$ and $P_{\Vert}$ are the interband matrix elements.
$m_{0}=9.10938188\times10^{-31}$ kg is the free electron mass. $\Delta(\mathbf{r})=\varepsilon_{g}\left(\mathbf{r}\right)/2$
is the gap energy parameter.

\begin{figure}
\includegraphics[width=8.5cm]{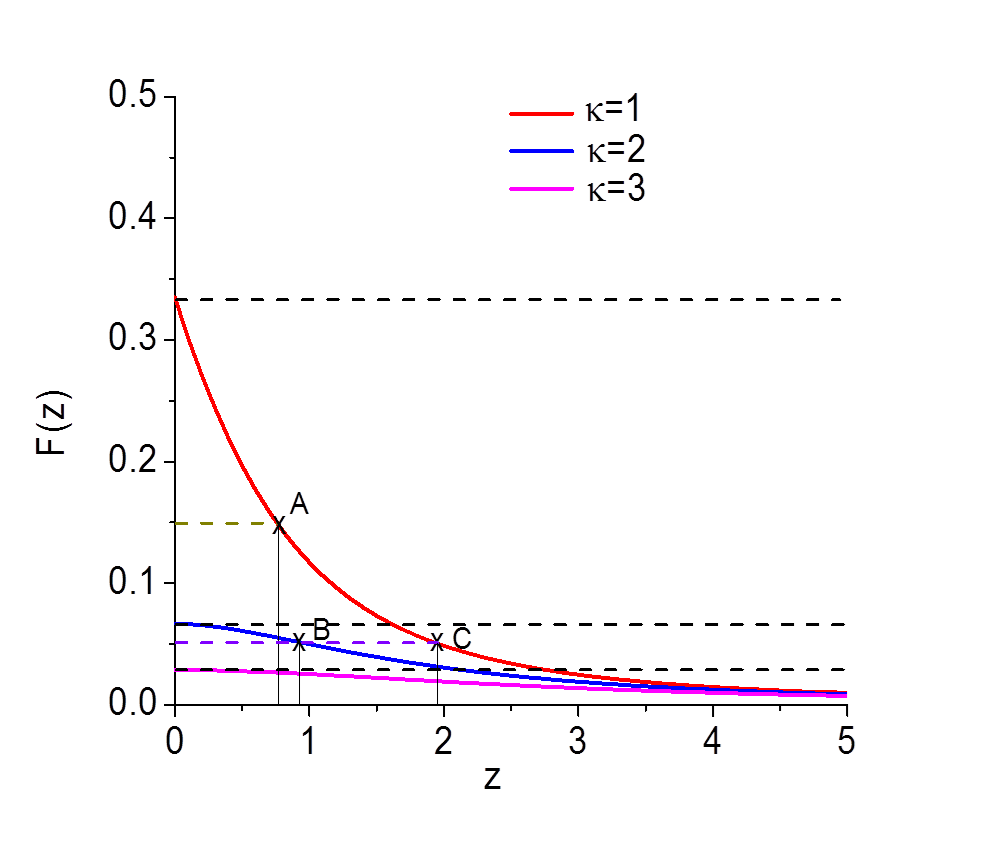}\caption{Plot of Eq.~(\ref{eq:7}) showing the intersections of the monotonically
decreasing $F\left(z\right)$ (solid lines) with the constants (black
dashed lines). Intersection at $z=0$ gives the minimum threshold
of the size of a QD to have two bound states, one positive and one
negative energy state, for a given confining potential. For a larger
QD, multiple bound states exist, corresponding to multiple intersection
points. The intersection points A, B and C are example points where
we evaluate the wavefunctions. The energy of the bound states are
determined by the relation $z=\lambda r_{o}$. \label{EquationPlot}}
\end{figure}

Assuming spherical symmetry for the 3D TI QD, $\Delta\left(r\right)$
depends on the radial coordinate only which breaks the crystal symmetry
in radial direction and has the symmetry $\Delta\left(r-r_{0}\right)=-\triangle\left(r_{0}-r\right)$,
where $r_{0}$ is the radius of the QD. Therefore, the angular parts
are separated from the radial part of the Dirac Hamiltonian (\ref{eq:1}).
Thus, we can follow the derivation of the solution for the central-force
problem of a hydrogen atom in relativistic quantum mechanics.\cite{Sakurai_advancedQM}
The eigenfunctions of $H$ are four-component spinors $\Phi=\left[\begin{array}{c}
\phi_{-}\\
\phi_{+}
\end{array}\right]=\left[\begin{array}{c}
f_{-}(r)\mathcal{\mathscr{Y}}_{jl_{-}}^{m_{j}}\\
if_{+}(r)\mathcal{\mathscr{Y}}_{jl_{+}}^{m_{j}}
\end{array}\right]$, where $f_{-}$and $f_{+}$ are the radial functions and $\mathcal{\mathscr{Y}}_{jl_{-}}^{m_{j}}$
and $\mathcal{\mathscr{Y}}_{jl_{+}}^{m_{j}}$ are the normalized spin-angular
functions corresponding to the $L_{-}$ and $L_{+}$ band, respectively,
such as in Pb$_{1-x}$Sn$_{x}$Te.  After eliminating the angular parts, the radial part of
the Dirac Hamiltonian (\ref{eq:1}) takes the form 
\begin{equation}
H=\left(\begin{array}{cc}
\Delta\left(r\right) & -v_{\Vert}\hbar\left(\frac{d}{dr}-\frac{\kappa}{r}\right)\\
v_{\Vert}\hbar\left(\frac{d}{dr}+\frac{\kappa}{r}\right) & -\Delta\left(r\right)
\end{array}\right)\label{eq:2}
\end{equation}
 where $v_{\Vert}=2.24\times10^{5}$ m/s for Pb$_{1-x}$Sn$_{x}$Te
and $\kappa=\pm\left(j+\frac{1}{2}\right)$ is a nonzero positive
or negative integer, $j$ being the total angular momentum quantum
number. For given $\kappa$, it is known from relativistic quantum
mechanics that the angular momenta $l_{-}$ and $l_{+}$ for $\phi_{-}$
and $\phi_{+}$ are determined by the relations $-\kappa=j\left(j+1\right)-l_{-}\left(l_{-}+1\right)+1/4$
and $\kappa=j\left(j+1\right)-l_{+}\left(l_{+}+1\right)+1/4$, respectively.
By solving $H^{2}\Phi=\varepsilon^{2}\Phi$, we obtain 
\begin{equation}
\left(r^{2}\frac{d^{2}}{dr^{2}}+2r\frac{d}{dr}\right)F_{\mp}-\left(\lambda^{2}r^{2}+\kappa\left(\kappa\pm1\right)\right)F_{\mp}=\beta r^{2}\frac{d\Delta}{dr}F_{\pm}\label{eq:3}
\end{equation}
 where $F_{\pm}=rf_{\pm}$, $\beta=1/v_{\Vert}\hbar$ and $\lambda=\beta\sqrt{\left(\Delta_{0}^{2}-\varepsilon^{2}\right)}$.
$\lambda$ behaves like a wave vector $\mathbf{k}$ whose allowed
quantized values determine the particle's energy levels. In a flat
geometry of a thin layer of a 3D TI, $\Delta\left(z\right)$ can be
chosen to be $\Delta\left(z\right)=\Delta\left(\infty\right)\tanh\left(z/l\right)$.\cite{Paudel&Leuenberger,Volkov&Paknratov}
We adopt a similar potential along the radial direction of the form
$\Delta(r^{'})=\Delta_{o}sgn(r^{'}-r_{o})$. Hence, the source term
in Eq.~(\ref{eq:3}) is ${\color{blue}{\color{black}\mathcal{F}_{\pm}\left(r^{'}\right)=2\Delta_{o}\beta F_{\pm}\left(r_{o}\right)r^{'^{2}}\delta\left(r^{'}-r_{o}\right)}}$.
Eqs.~(\ref{eq:3}) can be solved by using the corresponding differential
equation for the Green's function, i.e. 
\begin{equation}
\left[\frac{d}{dr}\left(r^{2}\frac{d}{dr}\right)-\left(\lambda^{2}r^{2}+\kappa\left(\kappa\pm1\right)\right)\right]G_{\mp}=\delta\left(r-r^{'}\right)\label{eq:4}
\end{equation}
 The solutions regular at $r=0$ with outgoing wave behavior at $r\rightarrow\infty$
are the product of spherical modified Bessel functions of the order
$\kappa$ for $G_{-}$ and of the order $\kappa-1$ for $G_{+}$,
i.e. $G_{-}\left(r,\: r^{'},\:\lambda\right)=C_{-}\mathcal{I}_{\kappa}\left(\lambda r_{<}\right)\mathcal{K}_{\kappa}\left(\lambda r_{>}\right)$,
$G_{+}\left(r,\: r^{'},\:\lambda\right)=C_{+}\mathcal{I}_{\kappa-1}\left(\lambda r_{<}\right)\mathcal{K}_{\kappa-1}\left(\lambda r_{>}\right)$,
where $r_{<}$ $\left(r_{>}\right)$ is the smaller (larger) of $r$
and $r^{'}$. The functions $\mathcal{I}\left(\lambda r\right)$ and
$\mathcal{K}\left(\lambda r\right)$ are, respectively, the first
and the second kind of modified spherical Bessel functions, and $C_{\mp}$
are the normalization constants. These constants are determined by
the discontinuity in slope implied by the delta function in Eq.~(\ref{eq:4}).
Integration is performed at the interface of the QD along the radial
direction: $\left[r^{2}\frac{dG_{\mp}}{dr}\right]_{r^{'}-\eta}^{r^{'}+\eta}=1$,
where $\eta$ is an infinitesimal quantity with $\eta>0$. For $r=r^{'}+\eta$,
$r_{>}=r$, $r_{<}=r^{'}$ and for $r=r^{'}-\eta$, $r_{>}=r^{'}$,
$r_{<}=r$. Consequently, the normalization constants are $C_{-}=1/\lambda r_{0}^{'^{2}}W_{\kappa}$
and $C_{+}=1/\lambda r_{o}^{'^{2}}W_{\kappa-1}$, where $W_{\kappa}=\left[\mathcal{I}_{\kappa}\left(\lambda r^{'}\right)\mathcal{K}_{\kappa}^{'}\left(\lambda r\right)-\mathcal{I}_{\kappa}^{'}\left(\lambda r\right)\mathcal{K}_{\kappa}\left(\lambda r^{'}\right)\right]_{r=r^{'}}$
and $W_{\kappa-1}=\left[\mathcal{I}_{\kappa-1}\left(\lambda r^{'}\right)\mathcal{K}_{\kappa-1}^{'}\left(\lambda r\right)-\mathcal{I}_{\kappa-1}^{'}\left(\lambda r\right)\mathcal{K}_{\kappa-1}\left(\lambda r^{'}\right)\right]_{r=r^{'}}$
are the Wronskians of $\mathcal{I}\left(\lambda r\right)$ and $\mathcal{K}\left(\lambda r\right)$,
respectively, for $\kappa$ and $\kappa-1$ order, and $\mathcal{I}^{'}\left(\lambda r\right)$
and $\mathcal{K}^{'}\left(\lambda r\right)$ are derivatives of the
Bessel functions. The Wronskian of two linearly independent functions
is proportional to $1/r^{2}$ for Sturm-Liouville type equations such
as Eq.~(\ref{eq:4}) (see the App.~\ref{sec:Wronskian}). The solutions
of Eqs.~(\ref{eq:3}) are $F_{\mp}=\int G_{\mp}\left(r,\: r^{'},\:\lambda\right)\mathcal{F}_{\pm}(r^{'})dr^{'}=2\Delta_{o}\beta\int G_{\mp}\left(r,\: r^{'},\:\lambda\right)F_{\pm}\left(r_{o}\right)r^{'^{2}}\delta\left(r^{'}-r_{o}\right)dr^{'}$,
i.e. 
\begin{eqnarray}
F_{-}\left(r\right) & = & 2\Delta_{o}\beta F_{+}\left(r_{o}\right)\mathcal{I}_{\kappa}\left(\lambda r_{<}\right)\mathcal{K}_{\kappa}\left(\lambda r_{>}\right)/\lambda W_{\kappa}\label{eq:5}\\
F_{+}\left(r\right) & = & 2\Delta_{o}\beta F_{-}\left(r_{o}\right)\mathcal{I}_{\kappa-1}\left(\lambda r_{<}\right)\mathcal{K}_{\kappa-1}\left(\lambda r_{>}\right)/\lambda W_{\kappa-1}\label{eq:6}
\end{eqnarray}
 where $r_{<}$ $\left(r_{>}\right)$ is now the smaller (larger)
of $r$ and $r_{0}$. A transcendental equation is obtained by solving
Eqs.~(\ref{eq:5}) and (\ref{eq:6}) and evaluating at $r=r_{0}$,
\begin{equation}
\left[z\mathcal{I}_{\kappa}\left(z\right)\mathcal{K}_{\kappa}\left(z\right)\right]\left[z\mathcal{I}_{\kappa-1}\left(z\right)\mathcal{K}_{\kappa-1}\left(z\right)\right]=1/4\Delta_{o}^{2}\beta^{2}r_{o}^{2}\label{eq:7}
\end{equation}
 where $z=\lambda r_{0}$. In Fig. \ref{EquationPlot}, we show the
plot of Eq.~(\ref{eq:7}) where the function $F\left(z\right)$ is
defined as $F\left(z\right)=\left[z\mathcal{I}_{\kappa}\left(z\right)\mathcal{K}_{\kappa}\left(z\right)\right]\left[z\mathcal{I}_{\kappa-1}\left(z\right)\mathcal{K}_{\kappa-1}\left(z\right)\right]$.

\section{\label{sec:Bound-States-of}Bound States of the Weyl Fermions}

Each term in the square bracket on the left hand side of Eq.~(\ref{eq:7})
is a monotonically decreasing function of $z$ (for $z$ > 0), with
maximum value of $1/(2\kappa+1)$ for $\kappa^{th}$ order term and
$1/(2\kappa-1)$ for $(\kappa-1)^{th}$ order term occurring at $z=0$
(see the App.~\ref{sec:Bessel}). Therefore, their product has a
maximum value of $1/(4\kappa^{2}-1)$ at $z=0$ and is equal to $1/4\Delta_{o}^{2}\beta^{2}r_{o}^{2}$.
Since $F\left(z\right)$ is a monotonically decreasing function, for
each $\kappa$, there is at most a single solution given by the intersection
of $F\left(z\right)$ with the constant $1/4\Delta_{o}^{2}\beta^{2}r_{o}^{2}$
(dashed line and solid line in Fig. \ref{EquationPlot}). The critical
limit for having a single solution is determined by the intersection
at the maximum value of $F\left(z\right)$, which occurs at $z=0$.
This means that there exists a single solution of Eq.~(\ref{eq:7})
for each $\kappa$ as long as the condition $1/4\Delta_{o}^{2}\beta^{2}r_{o}^{2}\leq1/(4\kappa^{2}-1)$
is satisfied. Fig. \ref{EquationPlot} shows the plot of the first
three different values of $\kappa$, $\kappa=1$ (red), $2$ (blue)
and $3$ (pink), each a monotonically decreasing line (solid line)
cut by a horizontal line (dashed line) at most one time. Since $\lambda=\beta\sqrt{\left(\Delta_{0}^{2}-\varepsilon^{2}\right)}$,
each single solution gives rise to two bound states with same magnitude
but opposite sign of energy, giving rise to the mirror symmetry in
the energy spectrum. Indeed, this makes sense since Weyl fermions
are massless at zero band gap with the linear dispersion relation.
Note that there is no radial quantum number because in general a Dirac
potential allows only for a single positive-energy and a single negative-energy
solution in radial direction.

As the size of the QD grows, it is filled with more and more bound
states (see Fig. \ref{EquationPlot}) where for smaller value of $F\left(z\right)$,
a horizontal dashed line makes multiple cuts at different values of
the energy (i.e. $z$) for different $\kappa$. For negative $\kappa$,
the solutions diverge at the origin and are therefore physically not
valid. This result has profound implications because the sign of $\kappa$
determines whether $\mathbf{j}$ is parallel or antiparallel to the
spin $\mathbf{s}$ (see Ref. 26). Since $\kappa$ is only allowed
to be positive, only one spin orientation with respect to $\mathbf{j}$
is permitted. This corresponds to the spin locking effect, which is
a hallmark of 3D TIs. 
 This allows us to write down the more specific form of the spin-angular
functions, i.e. 
\begin{eqnarray}
\mathcal{\mathscr{Y}}_{jl_{-}}^{m_{j}} & = & -\sqrt{\frac{l_{-}-m_{j}+\frac{1}{2}}{2l_{-}+1}}Y_{l_{-}}^{m_{j}-\frac{1}{2}}\left[\begin{array}{c}
1\\
0
\end{array}\right]\nonumber \\
 &  & +\sqrt{\frac{l_{-}+m_{j}+\frac{1}{2}}{2l_{-}+1}}Y_{l_{-}}^{m_{j}+\frac{1}{2}}\left[\begin{array}{c}
0\\
1
\end{array}\right]\\
\mathcal{\mathscr{Y}}_{jl_{+}}^{m_{j}} & = & \sqrt{\frac{l_{+}+m_{j}+\frac{1}{2}}{2l_{+}+1}}Y_{l_{+}}^{m_{j}-\frac{1}{2}}\left[\begin{array}{c}
1\\
0
\end{array}\right]\nonumber \\
 &  & +\sqrt{\frac{l_{+}-m_{j}+\frac{1}{2}}{2l_{+}+1}}Y_{l_{+}}^{m_{j}+\frac{1}{2}}\left[\begin{array}{c}
0\\
1
\end{array}\right]
\end{eqnarray}
 where $l_{-}=j+\frac{1}{2}$ and $l_{+}=j-\frac{1}{2}$.

The condition $1/4\Delta_{o}^{2}\beta^{2}r_{o}^{2}=1/(4\kappa^{2}-1)$
determines the lower limit of the size of the QD to hold two bound
interface states, a positive and a negative energy state, for a given
value of the confining potential strength. The critical QD size depends
on the Fermi velocities and band gaps of the 3D TI materials. In Pb$_{1-x}$Sn$_{x}$Te,
$\Delta_{o}=0.0935$ eV, half of the band gap of PbTe. Choosing $v_{\Vert}=2.24\times10^{5}$m/s,\cite{Volkov&Paknratov}
results in a critical QD size of $r_{0}=1.4$ nm for $\kappa=1$ at
$z=0$. Similarly for $\kappa=2$ at $z=0$, the critical QD size
for Pb$_{1-x}$Sn$_{x}$Te is $r_{0}=3$ nm. The energy of the bound
states are determined from $z=\lambda r_{o}$, which gives a very
shallow energy level of $\varepsilon=\pm\Delta_{o}$ for $z=0$.


\begin{table}
\begin{tabular}{|c|c|c||c|c|c|}
\hline 
\multicolumn{1}{|c}{} & \multicolumn{1}{c|}{$\phi_{-}$} &  & \multicolumn{1}{c}{} & \multicolumn{1}{c|}{$\phi_{+}$} & \tabularnewline
\hline 
$\kappa$  & $l_{-}$  & $j$  & $\kappa$  & $l_{+}$  & $j$\tabularnewline
\hline 
1  & 1  & 1/2  & 1  & 0  & 1/2\tabularnewline
\hline 
2  & 2  & 3/2  & 2  & 1  & 3/2\tabularnewline
\hline 
3  & 3  & 5/2  & 3  & 2  & 5/2\tabularnewline
\hline 
4  & 4  & 7/2  & 4  & 3  & 7/2\tabularnewline
\hline 
\end{tabular}

\caption{$\phi_{-}$ and $\phi_{+}$ components}

\label{Table 1} 
\end{table}

For a given value of $\kappa$, quantum numbers characterizing the
wavefunctions $\phi_{-}${} and $\phi_{+}${} can be determined.
For $\kappa=1,\:2,\:3$ and $4$, the possible combination of the
quantum numbers are shown{} in Table \ref{Table 1} for both spinors
$\phi_{-}${} and $\phi_{+}$. Here we observe that the $\phi_{-}$
component is characterized by the spin being antiparallel to its angular
momentum, whereas the $\phi_{+}$ component is characterized by the
spin being parallel to its angular momentum. We show now how to identify
the Kramers pairs. According to Kramers theorem, which applies to
a time-reversal invariant system, a spin $1/2$ state is at least
twofold degenerate on the surface of a 3D TI. Hence, we obtain the
following examples of Kramers pairs. For $\kappa=1$, the 4-spinor
state with $m_{\frac{1}{2}}=\frac{1}{2}$, 
\begin{eqnarray}
\Phi_{\frac{1}{2},\frac{1}{2}}^{\kappa=1} & = & \left[\begin{array}{c}
f_{-}(r)\mathcal{\mathscr{Y}}_{\frac{1}{2}1}^{\frac{1}{2}}\\
if_{+}(r)\mathcal{\mathscr{Y}}_{\frac{1}{2}0}^{\frac{1}{2}}
\end{array}\right]\\
 & = & \left[\begin{array}{c}
f_{-}(r)\left(-\sqrt{\frac{1}{3}}Y_{1}^{0}\left[\begin{array}{c}
1\\
0
\end{array}\right]+\sqrt{\frac{2}{3}}Y_{1}^{1}\left[\begin{array}{c}
0\\
1
\end{array}\right]\right)\\
if_{+}(r)Y_{0}^{0}\left[\begin{array}{c}
1\\
0
\end{array}\right]
\end{array}\right]\nonumber 
\end{eqnarray}
 has as Kramers partner the 4-spinor state with $m_{\frac{1}{2}}=-\frac{1}{2}$,
\begin{eqnarray}
\Phi_{\frac{1}{2},-\frac{1}{2}}^{\kappa=1} & = & \left[\begin{array}{c}
f_{-}(r)\mathcal{\mathscr{Y}}_{\frac{1}{2}1}^{-\frac{1}{2}}\\
if_{+}(r)\mathcal{\mathscr{Y}}_{\frac{1}{2}0}^{-\frac{1}{2}}
\end{array}\right]\\
 & = & \left[\begin{array}{c}
f_{-}(r)\left(-\sqrt{\frac{2}{3}}Y_{1}^{-1}\left[\begin{array}{c}
1\\
0
\end{array}\right]+\sqrt{\frac{1}{3}}Y_{1}^{0}\left[\begin{array}{c}
0\\
1
\end{array}\right]\right)\\
if_{+}(r)Y_{0}^{0}\left[\begin{array}{c}
0\\
1
\end{array}\right]
\end{array}\right]\nonumber 
\end{eqnarray}
 For $\kappa=2$, the 4-spinor state with $m_{\frac{3}{2}}=\frac{3}{2}$,
\begin{eqnarray}
\Phi_{\frac{3}{2},\frac{3}{2}}^{\kappa=2} & = & \left[\begin{array}{c}
f_{-}(r)\mathcal{\mathscr{Y}}_{\frac{3}{2}2}^{\frac{3}{2}}\\
if_{+}(r)\mathcal{\mathscr{Y}}_{\frac{3}{2}1}^{\frac{3}{2}}
\end{array}\right]\\
 & = & \left[\begin{array}{c}
f_{-}(r)\left(-\sqrt{\frac{1}{5}}Y_{2}^{1}\left[\begin{array}{c}
1\\
0
\end{array}\right]+\sqrt{\frac{4}{5}}Y_{2}^{2}\left[\begin{array}{c}
0\\
1
\end{array}\right]\right)\\
if_{+}(r)Y_{1}^{1}\left[\begin{array}{c}
1\\
0
\end{array}\right]
\end{array}\right]\nonumber 
\end{eqnarray}
 has as Kramers partner the 4-spinor with $m_{\frac{3}{2}}=-\frac{3}{2}$,
\begin{eqnarray}
\Phi_{\frac{3}{2},-\frac{3}{2}}^{\kappa=2} & = & \left[\begin{array}{c}
f_{-}(r)\mathcal{\mathscr{Y}}_{\frac{3}{2}2}^{-\frac{3}{2}}\\
if_{+}(r)\mathcal{\mathscr{Y}}_{\frac{3}{2}1}^{-\frac{3}{2}}
\end{array}\right]\\
 & = & \left[\begin{array}{c}
f_{-}(r)\left(-\sqrt{\frac{4}{5}}Y_{2}^{-2}\left[\begin{array}{c}
1\\
0
\end{array}\right]+\sqrt{\frac{1}{5}}Y_{2}^{-1}\left[\begin{array}{c}
0\\
1
\end{array}\right]\right)\\
if_{+}(r)Y_{1}^{-1}\left[\begin{array}{c}
0\\
1
\end{array}\right]
\end{array}\right]\nonumber 
\end{eqnarray}
 For $\kappa=2$, the 4-spinor state with $m_{\frac{3}{2}}=\frac{1}{2}$,
\begin{eqnarray}
\Phi_{\frac{3}{2},\frac{1}{2}}^{\kappa=2} & = & \left[\begin{array}{c}
f_{-}(r)\mathcal{\mathscr{Y}}_{\frac{3}{2}2}^{\frac{1}{2}}\\
if_{+}(r)\mathcal{\mathscr{Y}}_{\frac{3}{2}1}^{\frac{1}{2}}
\end{array}\right]\\
 & = & \left[\begin{array}{c}
f_{-}(r)\left(-\sqrt{\frac{2}{5}}Y_{2}^{0}\left[\begin{array}{c}
1\\
0
\end{array}\right]+\sqrt{\frac{3}{5}}Y_{2}^{1}\left[\begin{array}{c}
0\\
1
\end{array}\right]\right)\\
if_{+}(r)\left(\sqrt{\frac{2}{3}}Y_{1}^{0}\left[\begin{array}{c}
1\\
0
\end{array}\right]+\sqrt{\frac{1}{3}}Y_{1}^{1}\left[\begin{array}{c}
0\\
1
\end{array}\right]\right)
\end{array}\right]\nonumber 
\end{eqnarray}
 has as Kramers partner the 4-spinor with $m_{\frac{3}{2}}=-\frac{1}{2}$,
\begin{eqnarray}
\Phi_{\frac{3}{2},-\frac{1}{2}}^{\kappa=2} & = & \left[\begin{array}{c}
f_{-}(r)\mathcal{\mathscr{Y}}_{\frac{3}{2}2}^{-\frac{1}{2}}\\
if_{+}(r)\mathcal{\mathscr{Y}}_{\frac{3}{2}1}^{-\frac{1}{2}}
\end{array}\right]\\
 & = & \left[\begin{array}{c}
f_{-}(r)\left(-\sqrt{\frac{3}{5}}Y_{2}^{-1}\left[\begin{array}{c}
1\\
0
\end{array}\right]+\sqrt{\frac{2}{5}}Y_{2}^{0}\left[\begin{array}{c}
0\\
1
\end{array}\right]\right)\\
if_{+}(r)\left(\sqrt{\frac{1}{3}}Y_{1}^{-1}\left[\begin{array}{c}
1\\
0
\end{array}\right]+\sqrt{\frac{2}{3}}Y_{1}^{0}\left[\begin{array}{c}
0\\
1
\end{array}\right]\right)
\end{array}\right]\nonumber 
\end{eqnarray}
 In general, the number of Kramers pairs is determined by the spin
multiplicity for each $m_{j}$ value.

\begin{figure}
\includegraphics[width=8.5cm]{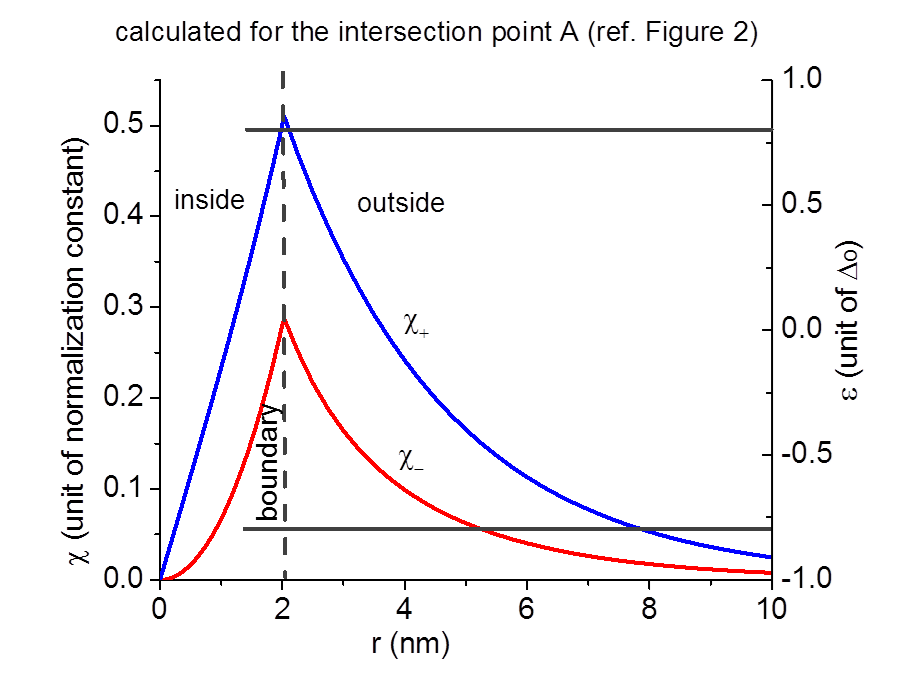}\caption{Spatial dependence of $f_{-}$ and $f_{+}$ inside and outside the
QD calculated for the intersection point A shown in Fig. \ref{EquationPlot}.
The QD has size $r_{0}=2$ nm. The solid horizontal lines represent
the energy eigenvalues $\varepsilon_{\pm}=\pm0.8\Delta_{o}$. \label{Figure 3}}
\end{figure}

\begin{figure}
\includegraphics[width=8.5cm]{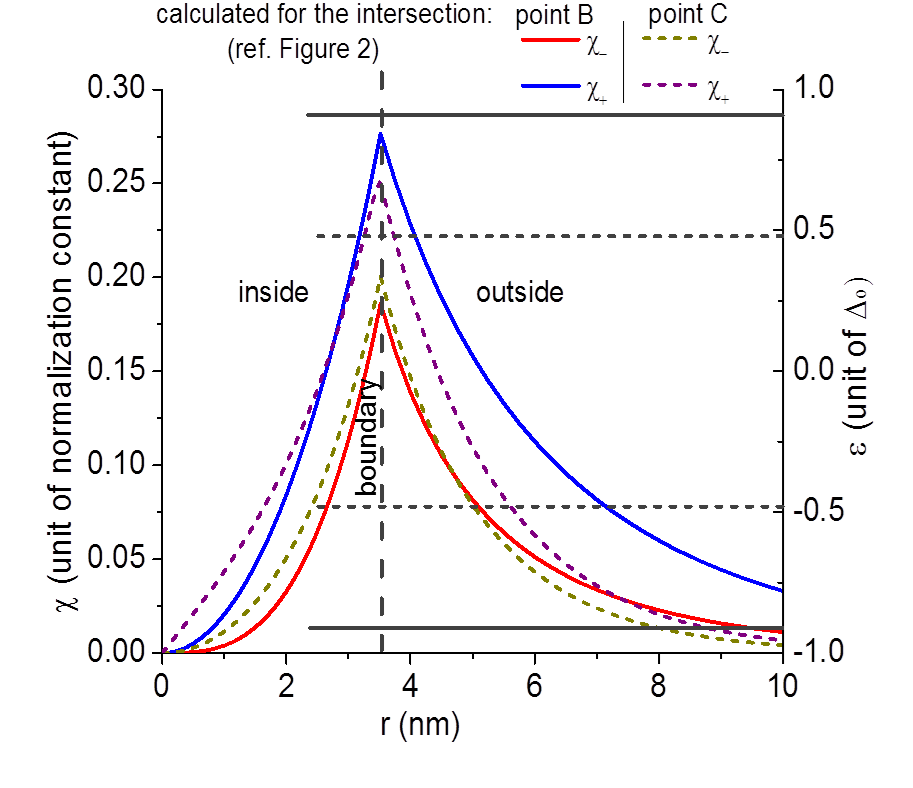}\caption{Spatial dependence of $f_{-}$ and $f_{+}$ inside and outside the
QD calculated for the intersection points B and C shown in Fig. \ref{EquationPlot}.
The QD has size $r_{0}=3.5$ nm. The horizontal solid lines represent
the energy eigenvalues $\varepsilon_{\pm}=\pm0.91\Delta_{o}$ at point
B and the dotted lines represents the energy eigenvalues $\varepsilon_{\pm}=\pm0.48\Delta_{o}$
at point C. \label{Figure 4} }
\end{figure}

In Figs. \ref{Figure 3} and \ref{Figure 4} we show the spatial wavefunctions
of the $f_{-}$ and $f_{+}$ components inside and outside the QD
made of the core-bulk heterostructure PbTe/Pb$_{0.31}$Sn$_{0.69}$Te.
The Fig. \ref{Figure 3} shows the example of the intersection point
A (see Fig. \ref{EquationPlot}) and the Fig. \ref{Figure 4} shows
the example of the intersection points B and C (see Fig. \ref{EquationPlot}).
Since the 4-spinors must be continuous at the boundary, also each
of the 2-spinor components must be continuous, i.e. $f_{-}^{in}=f_{-}^{out}$
and $f_{+}^{in}=f_{+}^{out}$ at the QD surface. The horizontal solid
and short dashed lines in Figs. represent the energy eigenvalues,
respectively, at the intersection point A, corresponding to $r_{0}=2$
nm, and at the intersection point B and C, corresponding to $r_{0}=3.5$
nm. Eigenvalues are $\varepsilon_{\pm}=\pm0.80\Delta_{o}$ at point
A, $\varepsilon_{\pm}=\pm0.91\Delta_{o}$ at point B, and $\varepsilon_{\pm}=\pm0.48\Delta_{o}$
at point C.

In order to show that the solutions correspond to Weyl fermions, we
perform an expansion of Eq.~(\ref{eq:7}) for large $z$ to obtain the eigenenergies in the continuum limit. Using the
second order in the expansion of the spherical modified Bessel functions
for $z\rightarrow\infty$ (see the App.~\ref{sec:Bessel}), we get

\begin{equation}
\frac{1}{2z}\left[1-\frac{2\kappa\left(\kappa+1\right)}{\left(2z\right)^{2}}\right]\times\frac{1}{2z}\left[1-\frac{2\kappa\left(\kappa-1\right)}{\left(2z\right)^{2}}\right]=1/4\Delta_{o}^{2}\beta^{2}r_{o}^{2}.\label{eq:8}
\end{equation}
 This can be written as
\begin{equation}
\varepsilon^{4}-\varepsilon^{2}\Delta_{o}^{2}+\frac{\Delta_{o}^{2}\kappa^{2}}{\beta^{2}r_{o}^{2}}=0\label{eq:9-1}
\end{equation}
 which results in the eigenenergies for the electron and hole,
\begin{equation}
\varepsilon_{\pm}=\pm\kappa v_{\shortparallel}\hbar/r_{o}.
\label{eq:10}
\end{equation}
This corresponds to the linear spectrum of free massless Dirac fermions,
i.e. free Weyl fermions on a sphere. This means that the energy splittings
between the trapped Weyl states in the quantum dot result from the
confinement of the Weyl fermions on a sphere. The solution
in Eq. (\ref{eq:10}) corresponds to the eigenspectra found in Ref.~\onlinecite{Lee:2009}
for zero magnetic field and without quantum confinement effects.

In the continuum limit, the Nielsen-Ninomiya
fermion doubling theorem \cite{Nielsen&Ninomiya} is satisfied by
the pairs of Dirac cones positioned at antipodal points of the sphere
defined by the surface of the QD (see App.~\ref{sec:doubling} for
details).
However, for a general finite QD radius $r_o$ the eigenstates are bound and have a discrete energy spectrum.
Since the Nielsen-Ninomiya fermion doubling theorem \cite{Nielsen&Ninomiya} is valid only for continuum states,
it does not apply to the bound Weyl fermions in a 3D TI QD with finite radius $r_o$.

\section{\label{sec:Optical-Excitations} Optical Excitations}

The $\mathbf{k}\cdot\mathbf{p}$ Hamiltonian contains also a quadratic
term in the momenta,\cite{Nimtz:1983} namely 
\begin{equation}
H_{q}=\left(\begin{array}{cc}
\frac{\left(p_{z}+eA_{z}\right)^{2}}{2m_{\Vert}^{-}}+\frac{\left(\mathbf{p}_{\bot}+e\mathbf{A}_{\bot}\right)^{2}}{2m_{\bot}^{-}} & 0\\
0 & \frac{\left(p_{z}+eA_{z}\right)^{2}}{2m_{\Vert}^{+}}+\frac{\left(\mathbf{p}_{\bot}+e\mathbf{A}_{\bot}\right)^{2}}{2m_{\bot}^{+}}
\end{array}\right),
\end{equation}
 where $m_{\Vert}^{\mp}$ and $m_{\bot}^{\mp}$ are the longitudinal
and transverse effective masses of the $L^{\mp}$ bands, respectively.
Through minimal coupling the quadratic term leads to a linear term
in the momentum, which we need to take into account. Hence, in the
presence of electromagnetic radiation, the total Hamiltonian for the
Dirac particle is given by \begin{widetext} 
\begin{eqnarray}
H_{{\rm tot}} & = & v_{\Vert}\alpha_{z}\left(\hat{p}_{z}+eA_{z}\right)+v_{\bot}\boldsymbol{\alpha}_{\bot}\cdot\left(\mathbf{\hat{p}}+e\mathbf{A}_{\perp}\right)+\beta\Delta-e\mathbf{\hat{r}}\cdot\mathbf{E}\nonumber \\
 & = & \left(\begin{array}{cc}
\Delta-e\mathbf{\hat{r}}\cdot\mathbf{E} & v_{\Vert}\sigma_{z}\left(\hat{p}_{z}+eA_{z}\right)+v_{\bot}\mathbf{\mathbf{\mathbf{\boldsymbol{\sigma}_{\bot}}}}\cdot\left(\mathbf{\hat{p}}+e\mathbf{A}_{\perp}\right)\\
v_{\Vert}\sigma_{z}\left(\hat{p}_{z}+eA_{z}\right)+v_{\bot}\mathbf{\mathbf{\mathbf{\boldsymbol{\sigma}_{\bot}}}}\cdot\left(\mathbf{\hat{p}}+e\mathbf{A}_{\perp}\right) & -\Delta-e\mathbf{\hat{r}}\cdot\mathbf{E}
\end{array}\right).
\end{eqnarray}
 \end{widetext} where $A=(A_{z},\;\boldsymbol{A}_{\perp})$ is the
vector potential, $\mathbf{E}=\partial\mathbf{A}/\partial t$ in the
Coulomb gauge, and we made use of the equivalence between $(e/m)\mathbf{A}\cdot\mathbf{p}$
and $-e\mathbf{\hat{r}}\cdot\mathbf{E}$.\cite{Sakurai_advancedQM}
We identify the interaction Hamiltonian as 
\begin{align}
 & H_{int}=ev_{\Vert}\alpha_{z}A_{z}+ev_{\bot}\boldsymbol{\alpha}_{\bot}\cdot\boldsymbol{A}_{\bot}-e\mathbf{\hat{r}}\cdot\mathbf{E}\\
= & \left(\begin{array}{cc}
-e\mathbf{\hat{r}}\cdot\mathbf{E} & ev_{\Vert}\sigma_{z}A_{z}+ev_{\bot}\mathbf{\mathbf{\mathbf{\boldsymbol{\sigma}_{\bot}}}}\cdot\mathbf{A}_{\perp}\\
ev_{\Vert}\sigma_{z}A_{z}+ev_{\bot}\mathbf{\mathbf{\mathbf{\boldsymbol{\sigma}_{\bot}}}}\cdot\mathbf{A}_{\perp} & -e\mathbf{\hat{r}}\cdot\mathbf{E}
\end{array}\right).\nonumber 
\end{align}
 It will turn out that both interband and intraband transitions contribute.
It is important to note that $v_{\Vert}=P_{\Vert}/m_{0}$ and $v_{\bot}=P_{\bot}/m_{0}$
include the Kane interband matrix elements $\mathbf{P}=\left\langle u_{\mathbf{k}_{\mathrm{f}}}^{\mp}\left|\mathbf{\hat{P}}\right|u_{\mathbf{k}_{\mathrm{I}}}^{\pm}\right\rangle $,
where $u_{\mathbf{k}}^{\mp}$ are the Bloch's functions for the $L^{\mp}$
bands. This means that the interband transitions are governed by the
interband Hamiltonian $H_{inter}=ev_{\Vert}\alpha_{z}A_{z}+ev_{\bot}\boldsymbol{\alpha}_{\bot}\cdot\boldsymbol{A}_{\bot}$,
where the Dirac $\boldsymbol{\alpha}$- matrices couple the $L^{-}$
band with the $L^{+}$ band. The Hamiltonian $H_{intra}=-e\mathbf{\hat{r}}\cdot\mathbf{E}$
accounts for intraband transitions with $\mathbf{\hat{r}}$ operating
on the envelope wavefunctions only. $H_{intra}$ is proportional to
the identity in 4-spinor space and therefore couples the $L^{-}$
band to itself and the $L^{+}$ band to itself. Thus the interband
Hamiltonian $H_{inter}$ and the intraband Hamiltonian $H_{intra}$
are not equivalent in this description. On the one hand, $H_{inter}$
gives rise to interband transitions because it contains the Kane interband
matrix elements $P_{\bot}$ and $P_{\Vert}$. On the other hand, $H_{intra}$
gives rise to intraband transitions because the electric dipole operator
$e\mathbf{\hat{r}}$ operates on the envelope wavefunctions. Remarkably,
both terms lead to the same strict optical selection rules and add
up to a combined optical matrix element, as shown below. This enhancement
of the optical matrix element is a feature of the 3D TI QD. In contrast,
in a wide-bandgap semiconductor QD the interband and intraband transitions
are energetically separated, i.e. interband transitions occur typically
around the bandgap energy, whereas intraband transitions occur around
the energy level separation due to the confinement of the QD. \cite{Singh}

\begin{figure}
\includegraphics[width=8.5cm]{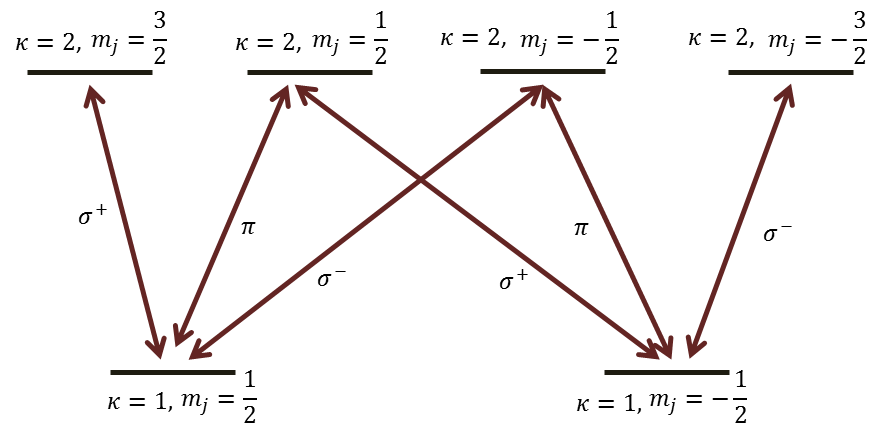}\caption{Optical transitions between the states $\kappa=1$ and $\kappa=2$
in 3D TI QD. Transitions are vertical. The transitions between $\left|\Phi_{\frac{1}{2},\pm\frac{1}{2}}^{\kappa=1}\right\rangle $
and $\left|\Phi_{\frac{3}{2},\pm\frac{1}{2}}^{\kappa=2}\right\rangle $
are coupled to the linear polarization of the incoming photons, while
the transitions between $\left|\Phi_{\frac{1}{2},\pm\frac{1}{2}}^{\kappa=1}\right\rangle $
and $\left|\Phi_{\frac{3}{2},\mp\frac{1}{2}}^{\kappa=2}\right\rangle $
and the transitions between $\left|\Phi_{\frac{1}{2},\pm\frac{1}{2}}^{\kappa=1}\right\rangle $
and $\left|\Phi_{\frac{3}{2},\pm\frac{3}{2}}^{\kappa=2}\right\rangle $
are coupled to the right and left circularly polarized light, respectively. }

\label{OpticalTransitions} 
\end{figure}

Fig. \ref{OpticalTransitions} shows the possible transitions between
the states $\kappa=1$ and $\kappa=2$. It is to be noted that there
is a complete symmetry in the solutions in the sense that a $\kappa$
state can be chosen from either the positive- or the negative-energy
solutions.. The optical matrix elements are given by

\begin{align}
\left\langle \phi_{f}\left|H_{int}\right|\phi_{I}\right\rangle = & ev_{\parallel}\left\langle \phi_{f}\left|\alpha_{z}\right|\phi_{I}\right\rangle A_{z}\nonumber \\
 & +ev_{\bot}\left\langle \phi_{f}\left|\boldsymbol{\alpha}_{\perp}\right|\phi_{I}\right\rangle \cdot\boldsymbol{A}_{\perp}\nonumber \\
 & -e\left\langle \phi_{f}\left|\mathbf{\hat{r}}\right|\phi_{I}\right\rangle \cdot\mathbf{E}\label{eq:9}
\end{align}
 The incoming photon's wavelength is much larger than the dot size.
Therefore, the transitions are vertical, which means $\mathbf{A}=(A_{x0},A_{y0},A_{z0})e^{i\mathbf{q}\cdot\mathbf{r}}\thickapprox(A_{x0},A_{y0},A_{z0})$
can be used, yielding the electric dipole approximation. The transition
energies $\hbar\omega_{0}=\varepsilon_{\kappa=2}-\varepsilon_{\kappa=1}$
are large compared with the room temperature $k_{B}T=25$ meV and
the Coulomb charging energy of about 5 meV.\cite{key-30} For the
control of the number of electrons and holes in the 3D TI QD it is
necessary to work at low temperatures of around 1 K.

As an example, here we consider transitions between the states $\kappa=1$
(at point C) and $\kappa=2$ (at point B). The matrix elements of
the Dirac-$\alpha$ matrix are given by 
\begin{equation}
\left\langle \Phi_{f}\left|\boldsymbol{\alpha}\right|\Phi_{I}\right\rangle =\left\langle \phi_{-}^{\kappa=2}\left|\boldsymbol{\sigma}\right|\phi_{+}^{\kappa=1}\right\rangle +\left\langle \phi_{+}^{\kappa=2}\left|\boldsymbol{\sigma}\right|\phi_{-}^{\kappa=1}\right\rangle 
\end{equation}
 The matrix elements of $\mathbf{\hat{r}}$ are given by 
\begin{equation}
\left\langle \phi_{f}\left|\mathbf{\hat{r}}\right|\phi_{I}\right\rangle =\left\langle \phi_{-}^{\kappa=2}\left|\mathbf{\hat{r}}\right|\phi_{-}^{\kappa=1}\right\rangle +\left\langle \phi_{+}^{\kappa=2}\left|\mathbf{\hat{r}}\right|\phi_{+}^{\kappa=1}\right\rangle 
\end{equation}
 The spherical harmonics can be determined using the Table \ref{Table 1}.
In order to obtain optical selection rules for circular polarizations,
it is useful to express the scalar products of the interband and the
intraband Hamiltonian in the form $\mathbf{e}\cdot\boldsymbol{\alpha}=\mathbf{e}_{z}\alpha_{z}+\mathbf{e}_{-}\alpha_{+}+\mathbf{e}_{+}\alpha_{-}$
and $\mathbf{e}\cdot\mathbf{\hat{r}}=\mathbf{e}_{z}\hat{z}+\mathbf{e}_{-}\hat{r}_{+}+\mathbf{e}_{+}\hat{r}_{-}$,
respectively, where $\mathbf{e}_{\pm}=\left(\mathbf{e}_{x}\pm i\mathbf{e}_{y}\right)/\sqrt{2}$
are the unit vectors of circular polarizations, $\alpha_{\pm}=\left(\alpha_{x}\pm i\alpha_{y}\right)/\sqrt{2}$,
and $\hat{r}_{\pm}=\left(\hat{x}\pm i\hat{y}\right)/\sqrt{2}=-\sqrt{\frac{4\pi}{3}}rY_{1}^{\pm1}$.
Using our spinor states $\left|\phi_{\mp}\right\rangle $ and radial
wavefunction functions $\left|f_{\mp}\right\rangle $ we obtain the
following nonzero matrix elements for $\boldsymbol{\alpha}$: 
\begin{eqnarray}
\left\langle \Phi_{\frac{3}{2},\pm\frac{1}{2}}^{\kappa=2}\left|\alpha_{z}\right|\Phi_{\frac{1}{2},\pm\frac{1}{2}}^{\kappa=1}\right\rangle  & \overset{\pi}{=} & \left\langle \phi_{+,\frac{3}{2},\pm\frac{1}{2}}^{\kappa=2}\left|\sigma_{z}\right|\phi_{-,\frac{1}{2},\pm\frac{1}{2}}^{\kappa=1}\right\rangle \nonumber \\
 & = & \frac{2\sqrt{2}}{3}i\left\langle f_{+}\mid f_{-}\right\rangle \\
\left\langle \Phi_{\frac{3}{2},-\frac{1}{2}}^{\kappa=2}\left|\alpha_{-}\right|\Phi_{\frac{1}{2},+\frac{1}{2}}^{\kappa=1}\right\rangle  & \overset{\sigma^{-}}{=} & \left\langle \phi_{+,\frac{3}{2},-\frac{1}{2}}^{\kappa=2}\left|\sigma_{-}\right|\phi_{-,\frac{1}{2},+\frac{1}{2}}^{\kappa=1}\right\rangle \nonumber \\
 & = & \frac{2}{3}i\left\langle f_{+}\mid f_{-}\right\rangle \\
\left\langle \Phi_{+,\frac{3}{2},+\frac{1}{2}}^{\kappa=2}\left|\alpha_{+}\right|\Phi_{-,\frac{1}{2},-\frac{1}{2}}^{\kappa=1}\right\rangle  & \overset{\sigma^{+}}{=} & \left\langle \phi_{+,\frac{3}{2},+\frac{1}{2}}^{\kappa=2}\left|\sigma_{+}\right|\phi_{-,\frac{1}{2},-\frac{1}{2}}^{\kappa=1}\right\rangle \nonumber \\
 & = & -\frac{2}{3}i\left\langle f_{+}\mid f_{-}\right\rangle \\
\left\langle \Phi_{\frac{3}{2},+\frac{3}{2}}^{\kappa=2}\left|\alpha_{+}\right|\Phi_{\frac{1}{2},+\frac{1}{2}}^{\kappa=1}\right\rangle  & \overset{\sigma^{+}}{=} & \left\langle \phi_{+,\frac{3}{2},+\frac{3}{2}}^{\kappa=2}\left|\sigma_{+}\right|\phi_{-,\frac{1}{2},+\frac{1}{2}}^{\kappa=1}\right\rangle \nonumber \\
 & = & -\frac{2}{\sqrt{3}}i\left\langle f_{+}\mid f_{-}\right\rangle \\
\left\langle \Phi_{+,\frac{3}{2},-\frac{3}{2}}^{\kappa=2}\left|\alpha_{-}\right|\Phi_{-,\frac{1}{2},-\frac{1}{2}}^{\kappa=1}\right\rangle  & \overset{\sigma^{-}}{=} & \left\langle \phi_{+,\frac{3}{2},-\frac{3}{2}}^{\kappa=2}\left|\sigma_{-}\right|\phi_{-,\frac{1}{2},-\frac{1}{2}}^{\kappa=1}\right\rangle \nonumber \\
 & = & \frac{2}{\sqrt{3}}i\left\langle f_{+}\mid f_{-}\right\rangle 
\end{eqnarray}
 For $\mathbf{\hat{r}}$ we obtain the following nonzero matrix elements:
\begin{eqnarray}
\left\langle \Phi_{\frac{3}{2},+\frac{1}{2}}^{\kappa=2}\left|\hat{z}\right|\Phi_{\frac{1}{2},+\frac{1}{2}}^{\kappa=1}\right\rangle  & \overset{\pi}{=} & \sqrt{\frac{2}{15}}\left\langle f_{-}Y_{2}^{0}\right|\hat{z}\left|f_{-}Y_{1}^{0}\right\rangle \nonumber \\
 &  & +\sqrt{\frac{6}{15}}\left\langle f_{-}Y_{2}^{1}\right|\hat{z}\left|f_{-}Y_{1}^{1}\right\rangle \nonumber \\
 &  & +\sqrt{\frac{2}{3}}\left\langle f_{+}Y_{1}^{0}\right|\hat{z}\left|f_{+}Y_{0}^{0}\right\rangle , \\
\left\langle \Phi_{\frac{3}{2},-\frac{1}{2}}^{\kappa=2}\left|\hat{z}\right|\Phi_{\frac{1}{2},-\frac{1}{2}}^{\kappa=1}\right\rangle  & \overset{\pi}{=} & \sqrt{\frac{6}{15}}\left\langle f_{-}Y_{2}^{-1}\right|\hat{z}\left|f_{-}Y_{1}^{-1}\right\rangle \nonumber \\
 &  & +\sqrt{\frac{2}{15}}\left\langle f_{-}Y_{2}^{0}\right|\hat{z}\left|f_{-}Y_{1}^{0}\right\rangle \nonumber \\
 &  & +\sqrt{\frac{2}{3}}\left\langle f_{+}Y_{1}^{0}\right|\hat{z}\left|f_{+}Y_{0}^{0}\right\rangle ,
\end{eqnarray}
\begin{eqnarray}
\left\langle \Phi_{\frac{3}{2},-\frac{1}{2}}^{\kappa=2}\left|\hat{r}_{-}\right|\Phi_{\frac{1}{2},+\frac{1}{2}}^{\kappa=1}\right\rangle  & \overset{\sigma^{-}}{=} & \sqrt{\frac{3}{15}}\left\langle f_{-}Y_{2}^{-1}\right|\hat{r}_{-}\left|f_{-}Y_{1}^{0}\right\rangle \nonumber \\
 &  & +\sqrt{\frac{4}{15}}\left\langle f_{-}Y_{2}^{0}\right|\hat{r}_{-}\left|f_{-}Y_{1}^{1}\right\rangle \nonumber \\
 &  & +\sqrt{\frac{1}{3}}\left\langle f_{+}Y_{1}^{-1}\right|\hat{r}_{-}\left|f_{+}Y_{0}^{0}\right\rangle ,  \\
\left\langle \Phi_{\frac{3}{2},+\frac{1}{2}}^{\kappa=2}\left|\hat{r}_{+}\right|\Phi_{\frac{1}{2},-\frac{1}{2}}^{\kappa=1}\right\rangle  & \overset{\sigma^{+}}{=} & \sqrt{\frac{4}{15}}\left\langle f_{-}Y_{2}^{0}\right|\hat{r}_{+}\left|f_{-}Y_{1}^{-1}\right\rangle \nonumber \\
 &  & +\sqrt{\frac{3}{15}}\left\langle f_{-}Y_{2}^{1}\right|\hat{r}_{+}\left|f_{-}Y_{1}^{0}\right\rangle \nonumber \\
 &  & +\sqrt{\frac{1}{3}}\left\langle f_{+}Y_{1}^{1}\right|\hat{r}_{+}\left|f_{+}Y_{0}^{0}\right\rangle , 
\end{eqnarray}
\begin{eqnarray}
\left\langle \Phi_{\frac{3}{2},+\frac{3}{2}}^{\kappa=2}\left|\hat{r}_{+}\right|\Phi_{\frac{1}{2},+\frac{1}{2}}^{\kappa=1}\right\rangle  & \overset{\sigma^{+}}{=} & \sqrt{\frac{1}{15}}\left\langle f_{-}Y_{2}^{1}\right|\hat{r}_{+}\left|f_{-}Y_{1}^{0}\right\rangle \nonumber \\
 &  & +\sqrt{\frac{8}{15}}\left\langle f_{-}Y_{2}^{2}\right|\hat{r}_{+}\left|f_{-}Y_{1}^{1}\right\rangle \nonumber \\
 &  & +\left\langle f_{+}Y_{1}^{1}\right|\hat{r}_{+}\left|f_{+}Y_{0}^{0}\right\rangle ,  \\
\left\langle \Phi_{\frac{3}{2},-\frac{3}{2}}^{\kappa=2}\left|\hat{r}_{-}\right|\Phi_{\frac{1}{2},-\frac{1}{2}}^{\kappa=1}\right\rangle  & \overset{\sigma^{-}}{=} & \sqrt{\frac{8}{15}}\left\langle f_{-}Y_{2}^{-2}\right|\hat{r}_{-}\left|f_{-}Y_{1}^{-1}\right\rangle \nonumber \\
 &  & +\sqrt{\frac{1}{15}}\left\langle f_{-}Y_{2}^{-1}\right|\hat{r}_{-}\left|f_{-}Y_{1}^{0}\right\rangle \nonumber \\
 &  & +\left\langle f_{+}Y_{1}^{-1}\right|\hat{r}_{-}\left|f_{+}Y_{0}^{0}\right\rangle ,
\end{eqnarray}
 where $\sigma_{+}=\left(\sigma_{x}+i\sigma_{y}\right)/\sqrt{2}=\left(\begin{array}{cc}
0 & \sqrt{2}\\
0 & 0
\end{array}\right)$, $\sigma_{-}=\left(\sigma_{x}-i\sigma_{y}\right)/\sqrt{2}=\left(\begin{array}{cc}
0 & 0\\
\sqrt{2} & 0
\end{array}\right)$, and the normalization and orthogonality condition $\underset{\Omega}{\int}d\Omega Y^{*m^{'},l^{'}}\left(\Omega\right)Y^{m,l}\left(\Omega\right)=\delta_{l^{'}l}\delta_{m^{'}m}$
have been used. All other matrix elements are zero.

The transition energy difference between the states $\kappa=1$ (at
point C) and $\kappa=2$ (at point B) is $0.43\Delta_{o}$ ($\lambda=7.5\:\mu m$)
and $1.39\Delta_{o}$ within the same energy solution and between
the negative and positive energy solutions, respectively (see Fig.
\ref{eq:4}). For $\Delta_{o}=93.5$ meV (half of the band gap of
PbTe), the corresponding wavelengths are 31$\:\mu m$ and 9.5$\:\mu m$.
Consider the transitions as shown in Fig. \ref{eq:5}. Using the Table
\ref{Table 1} to determine the spherical harmonics, we find that
the $z$-component of the matrix element gives rise to $\pi$-transitions
with $\kappa=2,\: m_{j}=1/2\longleftrightarrow\kappa=2,\: m_{j}=1/2$
and with $\kappa=2,\: m_{j}=-1/2\longleftrightarrow\kappa=1,\: m_{j}=-1/2$).
Thus, these $\pi$-transitions are coupled to light polarized linearly
in $z$-direction. The $x-iy$- and $x+iy$-components of the matrix
element give rise to the $\sigma^{+}$-transition with $\kappa=2,\: m_{j}=+1/2\longleftrightarrow\kappa=1,\: m_{j}=-1/2$
and with $\kappa=2,\: m_{j}=3/2\longleftrightarrow\kappa=1,\: m_{j}=1/2$
and to the $\sigma^{-}$-transition with $\kappa=2,\: m_{j}=-1/2\longleftrightarrow\kappa=1,\: m_{j}=+1/2$
and with $\kappa=2,\: m_{j}=-3/2\longleftrightarrow\kappa=1,\: m_{j}=-1/2$.
Thus, $\sigma^{+}$-transition and $\sigma^{-}$-transition are coupled
to the right and left circularly polarized light, respectively. We
can take advantage of these strict optical selection rules to implement
the semi-classical and quantum Faraday effect shown below. The overlap
integrals $\left\langle f_{+}\left(\kappa=2\right)\mid f_{-}\left(\kappa=1\right)\right\rangle $
and $\left\langle f_{-}\left(\kappa=2\right)\mid f_{+}\left(\kappa=1\right)\right\rangle $
for the transitions between the points B and C (in Fig. \ref{EquationPlot})
are evaluated to be 0.31 and 0.24, respectively. The Kane energy,
$E_{p}=2P_{\perp}^{2}/m_{o}$, is calculated to be 7.3 eV which is
about 3 times smaller than the Kane energy value of 22.7 eV for GaAs.\cite{Bastard,Rosencher}
The smaller Kane energy here is due to the fact that the Fermi velocity
is an order of magnitude smaller than the Fermi velocity in GaAs.
The polarization matrix elements of $\hat{r}_{\mp}$ accounts for
the strength of the in-plane intraband transitions at the band crossing.
We calculate the magnitude of the matrix elements for $\sigma^{\mp}$
transitions and find that $e\left|\left\langle \Phi_{\frac{3}{2},-\frac{1}{2}}^{\kappa=2}\left|\hat{r}_{-}\right|\Phi_{\frac{1}{2},+\frac{1}{2}}^{\kappa=1}\right\rangle \right|=e\left|\left\langle \Phi_{\frac{3}{2},+\frac{1}{2}}^{\kappa=2}\left|\hat{r}_{+}\right|\Phi_{\frac{1}{2},-\frac{1}{2}}^{\kappa=1}\right\rangle \right|=128$
Debye and $e\left|\left\langle \Phi_{\frac{3}{2},+\frac{3}{2}}^{\kappa=2}\left|\hat{r}_{+}\right|\Phi_{\frac{1}{2},+\frac{1}{2}}^{\kappa=1}\right\rangle \right|=e\left|\left\langle \Phi_{\frac{3}{2},-\frac{3}{2}}^{\kappa=2}\left|\hat{r}_{-}\right|\Phi_{\frac{1}{2},-\frac{1}{2}}^{\kappa=1}\right\rangle \right|=221$
Debye. For the $\pi$ transitions we find the magnitude of the matrix
elements as, $e\left|\left\langle \Phi_{\frac{3}{2},+\frac{1}{2}}^{\kappa=2}\left|\hat{z}\right|\Phi_{\frac{1}{2},+\frac{1}{2}}^{\kappa=1}\right\rangle \right|=e\left|\left\langle \Phi_{\frac{3}{2},-\frac{1}{2}}^{\kappa=2}\left|\hat{z}\right|\Phi_{\frac{1}{2},-\frac{1}{2}}^{\kappa=1}\right\rangle \right|=181$
Debye.

\section{\label{Faraday-Effect for QD}Faraday Effect for 3D TI QDs}

In Refs.~\onlinecite{Leuenberger:2005,Leuenberger:2006,Seigneur:2011,Gonzalez:2010,Seigneur:2010} we showed
that the single-photon Faraday rotation cannot only be used for quantum
spin memory but also for quantum teleportation and quantum computing
with wide-bandgap semiconductor QDs. In Ref.~\onlinecite{Thompson:2009} we showed that the conditional Faraday rotation can be used for optical switching of classical information. In Ref.~\onlinecite{Seigneur:2008} we proposed a single-photon Mach-Zehnder interferometer for quantum networks based on the single-photon Faraday effect. In Ref. \onlinecite{Berezovsky}
a single spin in a wide-bandgap semiconductor QD was detected using
the Faraday rotation. 
In order to implement these applications with 3D TI QDs, we need strict optical selection
rules for the circular polarization of the photons. Since, indeed, for 3D TI
QDs we obtain strict optical selection rules for circular polarization
of photons, we suggest that it is possible to implement quantum
memory, quantum teleportation, and quantum computing using the single-photon
Faraday rotation in 3D TI QDs. In order to prove this conjecture,
we derive the Faraday effect for 3D TI QDs. For the derivation
of the Faraday effect for a classical laser beam due to Pauli exclusion
principle we are going to follow Ref. \onlinecite{Davies}. Below in Sec.~\ref{Single-Photon-Faraday-Effect}
we are going to derive also the Faraday effect for a single photon
using quantum optical calculations, where we use Ref. \onlinecite{Scully&Zubairy}.

In order to simplify the notation, we write the light-matter interaction
Hamiltonian as $H_{int}=ev\boldsymbol{\alpha}\cdot\boldsymbol{A}-e\mathbf{\hat{r}}\cdot\mathbf{E}$.
Without loss of generality, the anisotropy coming from the band velocity
can be introduced back into the solutions at a later time. Since the
incident light is a plane wave with wavevector $\mathbf{q}$ and frequency
$\omega$ and the electric field component is $E=-\partial\boldsymbol{A}/\partial t$,
the interaction Hamiltonian reads 
\begin{eqnarray}
H_{int} & = & \frac{ePE_{0}}{im_{0}\omega}\left(e^{i\left(\mathbf{q\cdot r}-\omega t\right)}-e^{-i\left(\mathbf{q\cdot r}-\omega t\right)}\right)\boldsymbol{e\cdot\alpha}\nonumber \\
 &  & -eE_{0}\left(e^{i\left(\mathbf{q\cdot r}-\omega t\right)}+e^{-i\left(\mathbf{q\cdot r}-\omega t\right)}\right)\boldsymbol{e}\cdot\mathbf{\hat{r}}
\end{eqnarray}
 where $P=m_{0}v$ is the Kane interband matrix element. The transition
rate for a single 3D TI QD can then be calculated using Fermi's golden
rule, 
\begin{eqnarray}
W_{fI} & = & \frac{2\pi}{\hbar}\left(eE_{0}\right)^{2}\left|\left\langle \Phi_{f}\right|\frac{P}{im_{0}\omega}\boldsymbol{e\cdot\alpha}+\boldsymbol{e}\cdot\mathbf{\hat{r}}\left|\Phi_{I}\right\rangle \right|^{2}\nonumber \\
 &  & \times f\left(\varepsilon_{I}\right)\left[1-f\left(\varepsilon_{f}\right)\right]\delta\left(\varepsilon_{f}-\varepsilon_{I}\mp\hbar\omega\right)
\end{eqnarray}
 where $f\left(\varepsilon\right)=\left[\exp\left(\frac{\varepsilon-\varepsilon_{F}}{k_{B}T}\right)+1\right]$
is the Fermi-Dirac distribution function, $\varepsilon_{F}$ is the
Fermi energy, $\left|\Phi_{I}\right\rangle $ denotes the initial
Weyl state, $\left|\Phi_{f}\right\rangle $ denotes the final Weyl
state, and the - sign in front of $\hbar\omega$ corresponds to absorption
and the + sign to emission. Thus, the absorption of energy per spin
state is $\mathcal{P}=\hbar\omega\sum_{I,f}W_{fI}$. Comparing with
the total power $\mathcal{P}=2\sigma_{1}VE_{0}^{2}$ dissipated in
the system volume $V$, where $\sigma=\sigma_{1}+i\sigma_{2}$ is
the complex conductivity, and including absorption and emission, it
follows that the real part of the conductivity is 
\begin{eqnarray}
\sigma_{1} & = & \frac{\pi e^{2}\omega}{V}\sum_{I,f}\left|\left\langle \Phi_{f}\right|\frac{P}{im_{0}\omega}\boldsymbol{e\cdot\alpha}+\boldsymbol{e}\cdot\mathbf{\hat{r}}\left|\Phi_{I}\right\rangle \right|^{2}\nonumber \\
 &  & \times\left[f\left(\varepsilon_{I}\right)-f\left(\varepsilon_{f}\right)\right]\delta\left(\varepsilon_{f}-\varepsilon_{I}-\hbar\omega\right)
\end{eqnarray}
 which can be written in terms of the oscillator strengths $f_{fI}=\left(2m_{0}\omega_{fI}/\hbar\right)\left|\left\langle \Phi_{f}\right|\frac{P}{im_{0}\omega}\boldsymbol{e\cdot\alpha}+\boldsymbol{e}\cdot\mathbf{\hat{r}}\left|\Phi_{I}\right\rangle \right|^{2}$,
\begin{equation}
\sigma_{1}\left(\omega\right)=\frac{\pi e^{2}}{2m_{0}V}\sum_{fI}f_{fI}\left[f\left(\varepsilon_{I}\right)-f\left(\varepsilon_{f}\right)\right]\delta\left(\varepsilon_{f}-\varepsilon_{I}-\hbar\omega\right)
\end{equation}
 Using the relation $\epsilon_{r}=1+\frac{i}{\omega\epsilon_{0}}\sigma$,
where $\epsilon_{0}$ is the free-space permittivity, between the
complex conductivity and the complex dielectric function $\epsilon_{r}=\epsilon_{1}+i\epsilon_{2}$
and taking advantage of the Kramers-Kronig relations the complex dielectric
function is given by 
\begin{equation}
\epsilon_{r}\left(\omega\right)=1-\frac{e^{2}}{\epsilon_{0}m_{0}V}\sum_{fI}\frac{f_{fI}\left[f\left(\varepsilon_{I}\right)-f\left(\varepsilon_{f}\right)\right]}{\left(\omega^{2}-\omega_{fI}^{2}\right)+i\gamma\omega}
\end{equation}
 In order to describe the Faraday rotation, we need to consider only
the states $\left|\Phi_{\frac{1}{2},\pm\frac{1}{2}}^{\kappa=1}\right\rangle $,
$\left|\Phi_{\frac{3}{2},\pm\frac{1}{2}}^{\kappa=2}\right\rangle $,
and $\left|\Phi_{\frac{3}{2},\pm\frac{3}{2}}^{\kappa=2}\right\rangle $
coupled by circular polarized light (see Fig. \ref{OpticalTransitions}).
We denote their energy difference by $\hbar\omega_{0}=\varepsilon_{\kappa=2}-\varepsilon_{\kappa=1}$.
Defining the the quantity 
\begin{eqnarray}
M_{f;I} & = & \left|\left\langle \Phi_{f}^{\kappa=2}\right|\frac{P}{im_{0}\omega}\boldsymbol{e\cdot\alpha}+\boldsymbol{e}\cdot\mathbf{\hat{r}}\left|\Phi_{I}^{\kappa=1}\right\rangle \right|^{2}\nonumber \\
 &  & \times\left[f\left(\varepsilon_{I}\right)-f\left(\varepsilon_{f}\right)\right]
\end{eqnarray}
 we can rewrite the complex dielectric function as 
\begin{eqnarray}
\epsilon_{r}\left(\omega\right) & = & \epsilon_{QD}\left(\omega\right)-\frac{2e^{2}\rho\omega}{\epsilon_{0}\hbar}\left\{ \frac{M_{\frac{3}{2},+\frac{3}{2};\frac{1}{2},+\frac{1}{2}}+M_{\frac{3}{2},-\frac{3}{2};\frac{1}{2},-\frac{1}{2}}}{\left(\omega^{2}-\omega_{0}^{2}\right)+i\gamma\omega}\right.\nonumber \\
 &  & +\left.\frac{M_{\frac{3}{2},+\frac{1}{2};\frac{1}{2},-\frac{1}{2}}+M_{\frac{3}{2},-\frac{1}{2};\frac{1}{2},+\frac{1}{2}}}{\left[\omega^{2}-(\omega_{0}+\Delta_{S}/\hbar)^{2}\right]+i\gamma\omega}\right\} 
\end{eqnarray}
 where $\Delta_{S}$ is the Stark energy shift (see below). Summation
over the other states is included in $\epsilon_{QD}\left(\omega\right)$,
which is the dielectric function of Pb$_{0.63}$Sn$_{0.37}$Te, corresponding
to the material at the interface. $\rho=1/V$ is the 3D TI QD density.
This expression can be split into a component of the dielectric function
for the right circular polarization, 
\begin{eqnarray}
\epsilon_{+}\left(\omega\right) & = & \epsilon_{QD}\left(\omega\right)-\frac{2e^{2}\rho\omega}{\epsilon_{0}\hbar}\left\{ \frac{M_{\frac{3}{2},+\frac{3}{2};\frac{1}{2},+\frac{1}{2}}}{\left(\omega^{2}-\omega_{0}^{2}\right)+i\gamma\omega}\right.\nonumber \\
 &  & +\left.\frac{M_{\frac{3}{2},+\frac{1}{2};\frac{1}{2},-\frac{1}{2}}}{\left[\omega^{2}-(\omega_{0}+\Delta_{S}/\hbar)^{2}\right]+i\gamma\omega}\right\} 
\end{eqnarray}
 and a component of the dielectric function for the left circular
polarization, 
\begin{eqnarray}
\epsilon_{-}\left(\omega\right) & = & \epsilon_{QD}\left(\omega\right)-\frac{2e^{2}\rho\omega}{\epsilon_{0}\hbar}\left\{ \frac{M_{\frac{3}{2},-\frac{3}{2};\frac{1}{2},-\frac{1}{2}}}{\left(\omega^{2}-\omega_{0}^{2}\right)+i\gamma\omega}\right.\nonumber \\
 &  & +\left.\frac{M_{\frac{3}{2},-\frac{1}{2};\frac{1}{2},+\frac{1}{2}}}{\left[\omega^{2}-(\omega_{0}+\Delta_{S}/\hbar)^{2}\right]+i\gamma\omega}\right\} 
\end{eqnarray}
 Consequently, the indices of refraction for right and left circular
polarization are given by $n_{\pm}=\sqrt{\epsilon_{\pm}}$. Assuming
that the length of the material is $L$, the Faraday rotation can
now be understood by considering the electric component of the plane
wave after passing through the material at position $z=L$, 
\begin{eqnarray}
\mathbf{E}(z=L) & = & \frac{E_{0}}{\sqrt{2}}\left(e^{ik_{-}L}\mathbf{e}_{+}+e^{ik_{+}L}\mathbf{e}_{-}\right)e^{-i\omega t}\nonumber \\
 & = & E_{0}\left(\cos\frac{\Delta n\omega L}{c}\mathbf{e}_{x}+\sin\frac{\Delta n\omega L}{c}\mathbf{e}_{y}\right)\nonumber \\
 &  & \times e^{i\left(kL-\omega t+(n-1)\frac{\omega L}{c}\right)}
\end{eqnarray}
 where $\mathbf{e}_{\pm}=\left(\mathbf{e}_{x}\pm i\mathbf{e}_{y}\right)/\sqrt{2}$
are the circular polarization unit vectors, $n=\left(n_{+}+n_{-}\right)/2$
is the average index of refraction, $c$ is the speed of light in
vacuum, and $\Delta n=n_{+}-n_{-}$ is the difference in index of
refraction between right and left circular polarization. Thus, the
Faraday rotation angle is given by 
\begin{equation}
\vartheta=\frac{\Delta n\omega L}{2c}.\label{eq:20}
\end{equation}
 This formula shows that the Faraday rotation angle depends on the
populations of the states $\left|\Phi_{\frac{1}{2},\pm\frac{1}{2}}^{\kappa=1}\right\rangle $,
$\left|\Phi_{\frac{3}{2},\pm\frac{1}{2}}^{\kappa=2}\right\rangle $,
and $\left|\Phi_{\frac{3}{2},\pm\frac{3}{2}}^{\kappa=2}\right\rangle $,
as determined by the Fermi functions, which can be used in the quasi-equilibrium,
i.e. when the time is much smaller than the electron-hole recombination
time. A similar Faraday effect has already been successfully used
to experimentally detect a single spin inside a GaAs QD.\cite{Berezovsky}

\section{\label{sec:Quantum-Spin-Memory}Quantum Memory with 3D TI QDs}

\begin{figure}
\includegraphics[width=8.5cm]{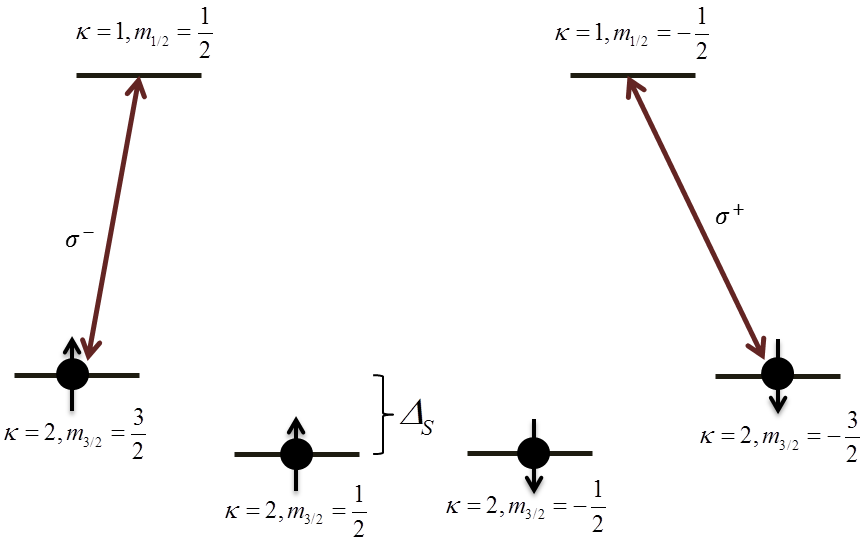}\caption{This is one possible level configuration that can be used for the
implementation of the quantum memory.}

\label{fig:QuantumMemory_configuration} 
\end{figure}

Let us first describe the quantum memory with 3D TI QDs. In order
to obtain the maximum Faraday effect, it is possible to apply an oscillating
electric field $\mathbf{E}(t)$ pointing in $z$-direction, which
splits the $\left|\Phi_{\frac{3}{2},\pm\frac{1}{2}}^{\kappa=2}\right\rangle $
states from the $\left|\Phi_{\frac{3}{2},\pm\frac{3}{2}}^{\kappa=2}\right\rangle $
states due to the optical Stark effect (see Fig. \ref{fig:QuantumMemory_configuration}).
The coupling to the electric field is described by the relativistic
Stark Hamiltonian 
\begin{equation}
H_{S}=\left(\begin{array}{cc}
-ezE_{z}e^{i\omega_{S}t} & ev_{\Vert}\sigma_{z}A_{z}\\
ev_{\Vert}\sigma_{z}A_{z} & -ezE_{z}e^{i\omega_{S}t}
\end{array}\right),
\end{equation}
 where $E_{z}(t)=E_{S}\left(e^{i\omega_{S}t}+e^{-i\omega_{S}t}\right)$
and thus $A_{z}(t)=\frac{iE_{S}}{\omega_{S}}\left(e^{i\omega_{S}t}-e^{-i\omega_{S}t}\right)$.
In second-order perturbation theory we obtain the quadratic Stark
effect. The only nonzero contributions come from the matrix element
coupling the $\left|\Phi_{\frac{3}{2},+\frac{1}{2}}^{\kappa=2}\right\rangle $
state to the $\left|\Phi_{\frac{1}{2},+\frac{1}{2}}^{\kappa=1}\right\rangle $
state, and from the matrix element coupling the$\left|\Phi_{\frac{3}{2},-\frac{1}{2}}^{\kappa=2}\right\rangle $
state to the $\left|\Phi_{\frac{1}{2},-\frac{1}{2}}^{\kappa=1}\right\rangle $
state. This yields the Stark energy shift 
\begin{eqnarray}
\Delta_{S} & = & e^{2}E_{S}^{2}\frac{\left|\left\langle \Phi_{\frac{1}{2},+\frac{1}{2}}^{\kappa=1}\right|\frac{P}{im_{0}\omega_{S}}\alpha_{z}+z\left|\Phi_{\frac{3}{2},+\frac{1}{2}}^{\kappa=2}\right\rangle \right|^{2}}{\hbar(\omega_{0}-\omega_{S})}\nonumber \\
 & = & e^{2}E_{S}^{2}\frac{\left|\left\langle \Phi_{\frac{1}{2},-\frac{1}{2}}^{\kappa=1}\right|\frac{P}{im_{0}\omega_{S}}\alpha_{z}+z\left|\Phi_{\frac{3}{2},-\frac{1}{2}}^{\kappa=2}\right\rangle \right|^{2}}{\hbar(\omega_{0}-\omega_{S})}\label{eq:33}
\end{eqnarray}

The Stark energy shift can be determined by applying an oscillating
electric field whose amplitude is measured along z-direction. The
amplitude of the electric field can be calculated as $\left|E_{S}\right|=\sqrt{2\mathit{\mathcal{S}}n/A\epsilon_{o}c}$,
where $\mathcal{S}$ is the power of the laser, $n$ is the index
of refraction of the medium through which the light propagates and
$A$ is the area of the aperture of the laser source. A laser power
of 1 mW with energy $\hbar\omega_{S}=30\: meV$ and area of the aperture
of $1\:\mu m^{2}$ in a medium with $n=5.7$ (for Pb$_{0.68}$Sn$_{0.32}$Te
at room temperature) can produce an electric field of $1.46\times10^{7}$
V/m. Using the Fermi velocity of $v_{\parallel}=2.24\times10^{5}$
m/s to calculate $P$, our calculations show that the matrix element
in Eq. (\ref{eq:33}) is $e\left|\left\langle \Phi_{\frac{1}{2},+\frac{1}{2}}^{\kappa=1}\right|\frac{P}{im_{0}\omega_{S}}\alpha_{z}+z\left|\Phi_{\frac{3}{2},+\frac{1}{2}}^{\kappa=2}\right\rangle \right|=410$
Debye. With the transition energy difference of $\hbar\omega_{0}=130$
meV we get a Stark energy shift of $\Delta_{S}=14$ meV.

\begin{figure}
\includegraphics[width=8.5cm]{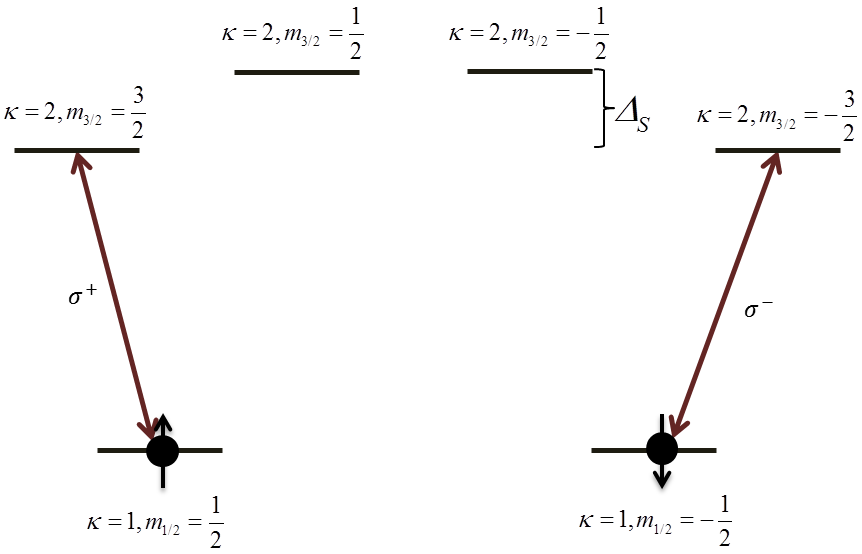}\caption{This is another possible level configuration that can be used for
the implementation of the quantum memory.}

\label{fig:QuantumMemory_configuration2} 
\end{figure}

It has already been shown experimentally that single-electron loading
is possible in 3D TI QDs. \cite{key-30} We focus on two possible
level configurations due to the electron-hole symmetry in 3D TI QDs: 
\begin{enumerate}
\item Fig. \ref{fig:QuantumMemory_configuration} shows the first level
configuration where the electron states are given by the s-like states
$\left|\Phi_{\frac{1}{2},\pm\frac{1}{2}}^{\kappa=1}\right\rangle $
and the hole states are given by the p-like states $\left|\Phi_{\frac{3}{2},\pm\frac{1}{2}}^{\kappa=2}\right\rangle $
and $\left|\Phi_{\frac{3}{2},\pm\frac{3}{2}}^{\kappa=2}\right\rangle $. 
\item Fig. \ref{fig:QuantumMemory_configuration2} shows the second level
configurationwhere the electron states are given by the p-like states
$\left|\Phi_{\frac{3}{2},\pm\frac{1}{2}}^{\kappa=2}\right\rangle $
and $\left|\Phi_{\frac{3}{2},\pm\frac{3}{2}}^{\kappa=2}\right\rangle $
and the hole states are given by the s-like states $\left|\Phi_{\frac{1}{2},\pm\frac{1}{2}}^{\kappa=1}\right\rangle $. 
\end{enumerate}
Only due to the symmetry between positive- and negative-energy solutions
in a 3D TI QD it is possible to choose either of these two level configurations.

Then, using the optical selection rules shown in Fig. \ref{OpticalTransitions},
we can use right circularly polarized light to create an e-h pair
with polarization +1, as shown in Fig. \ref{fig:QuantumMemory_configuration}.
This corresponds to writing the information +1 on the 3D TI QD. Alternatively,
we can use left circularly polarized light to create an e-h pair with
polarization -1, as shown in Fig. \ref{fig:QuantumMemory_configuration}.
This corresponds to writing the information -1 on the 3D TI QD.

\begin{figure}
\includegraphics[width=9cm]{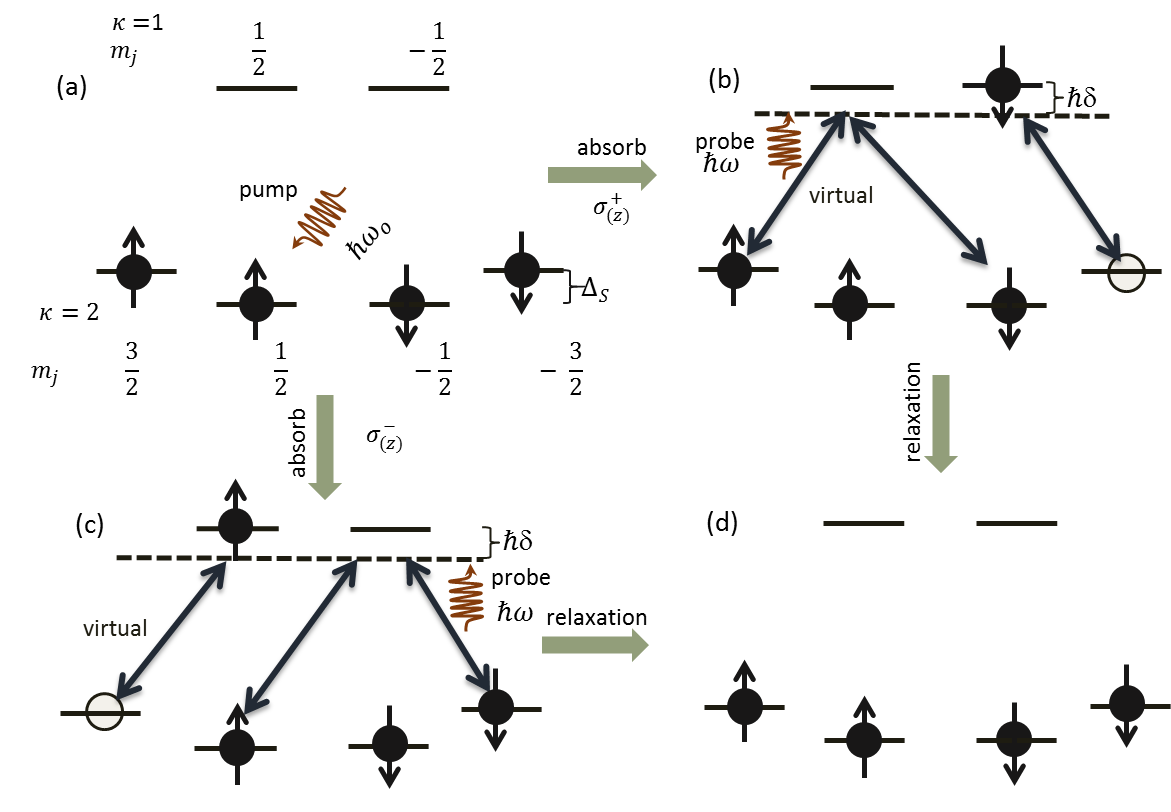}\caption{(a) The incident photon can be either right or left circularly polarized.
The initial level configuration is the one shown in Fig. \ref{fig:QuantumMemory_configuration}.
(b) If the photon is right circularly polarized, an e-h pair with
+1 polarization is created, which can be probed using off-resonant
linearly polarized light that acquires a negative Faraday rotation
angle through virtual excitation of an e-h pair and virtual recombinations
of the present e-h pair. (c) If the photon is left circularly polarized,
an e-h pair with -1 polarization is created, which can be probed using
off-resonant linearly polarized light that acquires a positive Faraday
rotation angle through virtual excitation of an e-h pair and virtual
recombinations of the present e-h pair. (d) After the probing, the
e-h pair relaxes into the ground state configuration.}

\label{fig:Faraday_rotation_exciton} 
\end{figure}

If we want to read out the information several times before the electron-hole
recombination, we can take advantage of the Faraday effect due to
the Pauli exclusion principle. For this method, we apply a $\pi$-pulse
of right or left circularly polarized light, thereby writing the information
+1 or -1, respectively, as shown in Fig. \ref{fig:Faraday_rotation_exciton}.
For +1 polarization, the Fermi functions, corresponding to populations
in quasi-equilibrium, are $f\left(\varepsilon_{\frac{1}{2},+\frac{1}{2}}\right)=0$,
$f\left(\varepsilon_{\frac{1}{2},-\frac{1}{2}}\right)=1$, $f\left(\varepsilon_{\frac{3}{2},+\frac{3}{2}}\right)=1$,
$f\left(\varepsilon_{\frac{3}{2},+\frac{1}{2}}\right)=1$, $f\left(\varepsilon_{\frac{3}{2},-\frac{1}{2}}\right)=1$
and $f\left(\varepsilon_{\frac{3}{2},-\frac{3}{2}}\right)=0$. For
-1 polarization, the Fermi functions, corresponding to populations
in quasi-equilibrium, are $f\left(\varepsilon_{\frac{1}{2},+\frac{1}{2}}\right)=1$,
$f\left(\varepsilon_{\frac{1}{2},-\frac{1}{2}}\right)=0$, $f\left(\varepsilon_{\frac{3}{2},+\frac{3}{2}}\right)=0$,
$f\left(\varepsilon_{\frac{3}{2},+\frac{1}{2}}\right)=1$ $f\left(\varepsilon_{\frac{3}{2},-\frac{1}{2}}\right)=1$
and $f\left(\varepsilon_{\frac{3}{2},-\frac{3}{2}}\right)=1$. Since
the off-resonant interaction does not destroy the quantum state on
the 3D TI QD, the information can be read out several times before
recombination. These results are in complete agreement with the quantum-optical
calculations shown below.

Let us assume a right circularly polarized pump pulse of energy $\hbar\omega_{0}$
excites an e-h pair with polarization +1 due to the $\sigma^{+}$-transition
from the state $\left|\Phi_{\frac{3}{2},-\frac{3}{2}}^{\kappa=2}\right\rangle $
to the state $\left|\Phi_{\frac{1}{2},-\frac{1}{2}}^{\kappa=1}\right\rangle $
in the level configuration shown in Fig. \ref{fig:QuantumMemory_configuration}.
Then a linearly polarized probe pulse of energy $\hbar\omega$ with
certain detuning energy is applied to read it out. There
are three virtual transitions that can occur while probing, one $\sigma^{-}$
transition: $\left|\Phi_{\frac{3}{2},\frac{3}{2}}^{\kappa=2}\right\rangle \longleftrightarrow\left|\Phi_{\frac{1}{2},+\frac{1}{2}}^{\kappa=1}\right\rangle $
and two $\sigma^{+}$ transitions: $\left|\Phi_{\frac{3}{2},-\frac{3}{2}}^{\kappa=1}\right\rangle \longleftrightarrow\left|\Phi_{\frac{1}{2},-\frac{1}{2}}^{\kappa=2}\right\rangle $ and $\left|\Phi_{\frac{3}{2},-\frac{1}{2}}^{\kappa=2}\right\rangle \longleftrightarrow\left|\Phi_{\frac{1}{2},+\frac{1}{2}}^{\kappa=1}\right\rangle$.
The matrix elements in the Eq. \ref{eq:20} are calculated using the
Fermi functions in the quasi-equilibrium. The matrix
elements are evaluated to be $M_{\frac{3}{2},\pm\frac{3}{2};\frac{1}{2},\pm\frac{1}{2}}=\mp8.07\times10^{-18}$
m$^{2}$ and $M_{\frac{3}{2},-\frac{1}{2};\frac{1}{2},+\frac{1}{2}}=-2.67\times10^{-18}$ m$^{2}$. 
The sign of the matrix elements $M_{f,I}$ is determined
by the Fermi functions. The corresponding dipole moments are
454 Debye and 261 Debye, respectively. For a quantitative estimate, we choose
a transition energy gap between the negative and positive energy solution
of $\hbar\omega_{o}=130\: meV$, a linearly polarized probe pulse
with detuning energy of $\hbar\delta=1$ meV and a cavity photon with
a bandwidth of $\hbar\gamma=100\:\mu$eV.\cite{Forchel,Berezovsky}
We further assume that there is a single QD in a slab material of
length $L=0.1\:\mu$m. With these values for our 3D
TI QD of size 3.5 nm we obtain the real part of the Faraday rotation
angle of $\vartheta_{+1}=-624\:\mu$rad. This Faraday rotation angle
is well above the angle value that has been measured for the experimental
detection of a single spin in GaAs QDs.\cite{Berezovsky} A similar
calculation can be done for a left circularly polarized pump pulse
that excites an e-h pair with polarization -1 due to a $\sigma^{-}$-transition
from the state $\left|\Phi_{\frac{3}{2},+\frac{3}{2}}^{\kappa=2}\right\rangle $
to the state $\left|\Phi_{\frac{1}{2},+\frac{1}{2}}^{\kappa=1}\right\rangle $.
Due to the symmetry of the positive- and negative-energy solutions
in 3D TI QDs, a large variety of level configurations can be considered
to achieve the Faraday effect.

The largest dipole moment of 452 Debye is one order of magnitude larger
than the typical value of 75 Debye for GaAs QDs,\cite{Stievater}
and two orders of magnitude larger than the typical value of a few
Debye for atoms.\cite{Takagahara} This large strength of the coupling
of infrared light to 3D TI QDs can partially compensate the weak overlap
of the photon with the 3D TI QD, which is due to the wavelength of
the infrared light being so much larger than the size of the 3D TI
QD.

\section{\label{Single-Photon-Faraday-Effect}Single-Photon Faraday Effect for
3D TI QDs}

Let us consider a 3D TI QD in the level configuration shown in Fig.
\ref{fig:QuantumMemory_configuration} inside a cavity. We define
$c_{1\pm}$, $c_{2\pm}$ and $c_{3\pm}$ as the annihilation operators
of the states $\left|\Phi_{\frac{3}{2},\pm\frac{1}{2}}^{\kappa=2}\right\rangle $,
$\left|\Phi_{\frac{3}{2},\pm\frac{3}{2}}^{\kappa=2}\right\rangle $,
and $\left|\Phi_{\frac{1}{2},\pm\frac{1}{2}}^{\kappa=1}\right\rangle $,
respectively. Then the Jaynes-Cummings model\cite{Scully&Zubairy}
gives rise to the Hamiltonian $H=H_{p}+H_{QD}+H_{p-QD}$, where 
\begin{eqnarray}
H_{p} & = & \hbar\omega_{c}\left(a_{+}^{\dagger}a_{+}+a_{-}^{\dagger}a_{-}\right),\\
H_{QD} & = & \sum_{j=1}^{3}\hbar\omega_{j}\left(c_{j+}^{\dagger}c_{j+}+c_{j-}^{\dagger}c_{j-}\right),\nonumber \\
H_{p-QD} & = & \hbar g_{1}\left(a_{+}c_{3+}^{\dagger}c_{1-}+a_{-}c_{3-}^{\dagger}c_{1+}\right)+h.c.\\
 &  & +\hbar g_{2}\left(a_{+}c_{3-}^{\dagger}c_{2-}+a_{-}c_{3+}^{\dagger}c_{2+}\right)+h.c.,
\end{eqnarray}
 are the cavity photon Hamiltonian, the QD Hamiltonian describing
the Weyl states, and the interaction Hamiltonian describing the photon-QD
interaction, respectively. We can safely neglect the vacuum energy
$\hbar\omega_{c}/2$ per mode. The photon-QD coupling constants are
given by $\hbar g_{1}=\sqrt{\hbar\omega/2\epsilon_{0}V_{0}}e\left\langle \Phi_{\frac{3}{2},\pm\frac{1}{2}}^{\kappa=2}\right|\frac{P}{im_{0}\omega}\boldsymbol{e\cdot\alpha}+\boldsymbol{e}\cdot\mathbf{\hat{r}}\left|\Phi_{\frac{1}{2},\mp\frac{1}{2}}^{\kappa=1}\right\rangle $
and $\hbar g_{2}=\sqrt{\hbar\omega/2\epsilon_{0}V_{0}}e\left\langle \Phi_{\frac{3}{2},\pm\frac{3}{2}}^{\kappa=2}\right|\frac{P}{im_{0}\omega}\boldsymbol{e\cdot\alpha}+\boldsymbol{e}\cdot\mathbf{\hat{r}}\left|\Phi_{\frac{1}{2},\pm\frac{1}{2}}^{\kappa=1}\right\rangle $,
where $V_{0}$ is the modal volume. After switching to the electron-hole
picture using the new electron and hole operators $c_{\pm}=c_{3\pm}$
and $v_{j\mp}^{\dagger}=c_{j\pm}$ for $j=1,2$, we obtain 
\begin{eqnarray}
H_{p} & = & \hbar\omega_{c}\left(a_{+}^{\dagger}a_{+}+a_{-}^{\dagger}a_{-}\right),\\
H_{QD} & = & \hbar\omega_{3}\left(c_{+}^{\dagger}c_{+}+c_{-}^{\dagger}c_{-}\right)+\sum_{j=1}^{2}\hbar\omega_{j}\left(v_{j+}^{\dagger}v_{j+}+v_{j-}^{\dagger}v_{j-}\right),\\
H_{int} & = & \hbar g_{1}\left(a_{+}c_{+}^{\dagger}v_{1+}^{\dagger}+a_{-}c_{-}^{\dagger}v_{1-}^{\dagger}\right)+h.c.\\
 &  & +\hbar g_{2}\left(a_{+}c_{-}^{\dagger}v_{2+}^{\dagger}+a_{-}c_{+}^{\dagger}v_{2-}^{\dagger}\right)+h.c.
\end{eqnarray}
where $\hbar\omega_{3}-\hbar\omega_{2}=\hbar\omega_{c}+\hbar\delta$
and $\hbar\omega_{2}-\hbar\omega_{1}=\Delta_{S}$. Since the interaction
between the EM fields and the QDs is off-resonant, we can apply
an adiabatic approximation. For that, let us calculate the time evolution
of the polarization operators $p_{j\sigma\sigma'}(t)=v_{j\sigma}c_{\sigma'}$
(coherences) by means of the Heisenberg equation of motion, i.e. 
\begin{equation}
\frac{\partial p_{j\sigma\sigma'}(t)}{\partial t}=\frac{1}{i\hbar}\left[p_{j\sigma\sigma'}(t),H\right]=\frac{1}{i\hbar}\left[p_{j\sigma\sigma'}(t),H_{{\rm QD}}+H_{{\rm int}}\right]
\end{equation}
 Since 
\begin{eqnarray}
\left[p_{j\sigma\sigma'},p_{\lambda j'\lambda'}^{\dagger}\right] & = & v_{j\sigma}v_{j'\lambda'}^{\dagger}\delta_{\sigma'\lambda}-c_{\lambda}^{\dagger}c_{\sigma'}\delta_{jj'}\delta_{\sigma\lambda'}\nonumber \\
 & = & \left(1-v_{j'\lambda'}^{\dagger}v_{j\sigma}\right)\delta_{\sigma'\lambda}-c_{\lambda}^{\dagger}c_{\sigma'}\delta_{jj'}\delta_{\sigma\lambda'}
\end{eqnarray}
 and
\begin{eqnarray}
\left[p_{j\sigma\sigma'},c_{\lambda}^{\dagger}c_{\lambda}\right] & = & p_{j\sigma\lambda}\delta_{\sigma'\lambda}\nonumber \\
\left[p_{j\sigma\sigma'},v_{j'\lambda}^{\dagger}v_{j'\lambda}\right] & = & p_{j'\lambda\sigma'}\delta_{jj'}\delta_{\sigma\lambda}
\end{eqnarray}
 we obtain
\begin{eqnarray}
i\frac{\partial p_{1\pm\pm}}{\partial t} & = & \left(\omega_{c}+\delta+\frac{\Delta_{S}}{\hbar}\right)p_{1\pm\pm} 
\nonumber\\
& & +g_{1}a_{\pm}\left(1-c_{\pm}^{\dagger}c_{\pm}-v_{1\pm}^{\dagger}v_{1\pm}\right) 
\nonumber\\
& & +g_{2}a_{\mp}\left(1-v_{2\mp}^{\dagger}v_{1\pm}\right),\\
i\frac{\partial p_{2\pm\mp}}{\partial t} & = & \left(\omega_{c}+\delta\right)p_{2\pm\mp}
\nonumber\\
& & +g_{1}a_{\mp}\left(1-v_{1\mp}^{\dagger}v_{2\pm}\right)
\nonumber\\
& & +g_{2}a_{\pm}\left(1-c_{\mp}^{\dagger}c_{\mp}-v_{2\pm}^{\dagger}v_{2\pm}\right),
\end{eqnarray}
where the + (-) sign denotes the polarization with right (left) circular
polarization. Since the states $\left|\Phi_{\frac{3}{2},\pm\frac{1}{2}}^{\kappa=2}\right\rangle $
and $\left|\Phi_{\frac{3}{2},\pm\frac{3}{2}}^{\kappa=2}\right\rangle $
are not resonantly coupled, no coherences $v_{2\pm}v_{1\mp}^{\dagger}$
and $v_{2\pm}v_{1\mp}^{\dagger}$ are created. Therefore they are
zero. It is possible to transform to the rotating frame by means of
$\widetilde{p}_{1\pm\pm}=p_{1\pm\pm}e^{-i\omega_{c}t}$, $\widetilde{p}_{2\pm\mp}=p_{2\pm\pm}e^{-i\omega_{c}t}$
and $\widetilde{a}_{\pm}=a_{\pm}e^{-i\omega_{c}t}$, resulting in
\begin{eqnarray}
i\frac{\partial p_{1\pm\pm}}{\partial t} & = & \left(\delta+\frac{\Delta_{S}}{\hbar}\right)p_{1\pm\pm} 
\nonumber\\
& & +g_{1}a_{\pm}\left(1-c_{\pm}^{\dagger}c_{\pm}-v_{1\pm}^{\dagger}v_{1\pm}\right),\\
i\frac{\partial p_{2\pm\mp}}{\partial t} & = & \delta p_{2\pm\mp}+g_{2}a_{\pm}\left(1-c_{\mp}^{\dagger}c_{\mp}-v_{2\pm}^{\dagger}v_{2\pm}\right),
\end{eqnarray}
 where we omitted the tildes. The Heisenberg equations for the polarization
operators $p_{1\pm\pm}^{\dagger}$ and $p_{2\pm\mp}^{\dagger}$ can
be obtained by taking the Hermitian conjugate. Since in the case of
the Faraday effect the photon is off-resonant with the energy difference
$\hbar\omega_{21}=\hbar\omega_{2}-\hbar\omega_{1}$, we can apply
the adiabatic approximation, which corresponds to setting the time
derivatives in the Heisenberg equations to zero, i.e. taking the stationary
limit. Then we obtain
\begin{eqnarray}
p_{1\pm\pm} & = & -\frac{g_{1}a_{\pm}\left(1-c_{\pm}^{\dagger}c_{\pm}-v_{1\pm}^{\dagger}v_{1\pm}\right)}{\left(\delta+\frac{\Delta_{S}}{\hbar}\right)},\\
p_{2\pm\mp} & = & -\frac{g_{2}a_{\pm}\left(1-c_{\mp}^{\dagger}c_{\mp}-v_{2\pm}^{\dagger}v_{2\pm}\right)}{\delta}.
\end{eqnarray}
 Inserting this result into the interaction Hamiltonian leads to an
effective interaction Hamiltonian of the form
\begin{eqnarray}
H_{int}^{eff} & = & -\frac{\hbar g_{1}^{2}}{\left(\delta+\frac{\Delta_{S}}{\hbar}\right)}\sum_{\sigma}\left(2a_{\sigma}^{\dagger}a_{\sigma}+1\right)\left(1-c_{\sigma}^{\dagger}c_{\sigma}-v_{1\sigma}^{\dagger}v_{1\sigma}\right)\nonumber \\
 &  & -\frac{\hbar g_{2}^{2}}{\delta}\sum_{\sigma}\left(2a_{\sigma}^{\dagger}a_{\sigma}+1\right)\left(1-c_{\bar{\sigma}}^{\dagger}c_{\bar{\sigma}}-v_{1\sigma}^{\dagger}v_{1\sigma}\right),
\end{eqnarray}
where $\bar{\sigma}$ has the opposite sign of $\sigma$. It becomes
obvious that if electrons or holes are present, the effective interaction
can be suppressed. Most importantly, this suppression of interaction
depends on the spin of the present electrons or holes. This is exactly
the mechanism for the Faraday effect due to Pauli exclusion principle.
Let us now calculate the time evolution of the photon operator in
the rotating frame under the effective interaction Hamiltonian, i.e.
\begin{eqnarray}
i\hbar\frac{\partial a_{\pm}}{\partial t} & = & \left[a_{\pm},H_{int}^{eff}\right] 
\nonumber \\
 & = & -2a_{\pm}\left[\left(\frac{\hbar g_{1}^{2}}{\delta+\frac{\Delta_{S}}{\hbar}}\right)\left(1-c_{\pm}^{\dagger}c_{\pm}-v_{1\pm}^{\dagger}v_{1\pm}\right)\right.
\nonumber\\
 &  & \left.+\left(\frac{\hbar g_{2}^{2}}{\delta}\right)\left(1-c_{\mp}^{\dagger}c_{\mp}-v_{2\pm}^{\dagger}v_{2\pm}\right)\right],
\end{eqnarray}
 resulting in the solution 
\begin{eqnarray}
a_{\pm}(t) & = & a_{\pm}(0)\exp\left\{-i\left[\frac{2g_{1}^{2}}{\delta+\frac{\Delta_{S}}{\hbar}}\left(1-c_{\pm}^{\dagger}c_{\pm}-v_{1\pm}^{\dagger}v_{1\pm}\right) \right.\right.
\nonumber\\
& & +\left.\left.\frac{2g_{2}^{2}}{\delta}\left(1-c_{\mp}^{\dagger}c_{\mp}-v_{2\pm}^{\dagger}v_{2\pm}\right)\right]t
\right\}.
\end{eqnarray}
 This formula is the main result of this section. It shows that the
Faraday rotation of the linearly polarized light depends strongly
on the presence of electrons and holes due to the Pauli exclusion
principle.

\section{\label{Quantum-Teleportation}Quantum Teleportation and Quantum Computing
with 3D TI QDs}

Here we show that the single-photon Faraday rotation can be used to
entangle a single photon with either a single e-h pair, a single electron,
or a single hole. This entanglement can be used as a resource to implement
optically mediated quantum teleportation and quantum computing 3D
TI QDs based on the Faraday effect due to the Pauli exclusion principle,
where the qubit is defined as either the polarization of a single
e-h pair, the spin of a single electron, or the spin of a single hole.
The quantum-informational methods for the implementation of quantum
teleportation and quantum computing are described in Refs. \onlinecite{Leuenberger:2005,Leuenberger:2006}.
We describe here the physical methods for creating the entanglement.

\subsection{Photon polarization - e-h pair polarization entanglement}

Let us consider now the Faraday effect due to an e-h pair on the QD
for the level configuration shown in Fig. \ref{fig:QuantumMemory_configuration2}.
The initial state before the photon-QD interaction reads 
\begin{equation}
\left|\psi_{\mp1}(0)\right\rangle =\frac{1}{\sqrt{2}}\left(e^{-i\vartheta_{0}}a_{+}^{\dagger}+e^{i\vartheta_{0}}a_{-}^{\dagger}\right)c_{\pm}^{\dagger}v_{2\mp}^{\dagger}\left|0\right\rangle 
\end{equation}
 where the photon is linearly polarized at an angle $\vartheta_{0}$
from the $x$-axis. If the initial e-h pair is $-1$ polarized, then
the state after time $t$ is given by 
\begin{eqnarray}
\left|\psi_{-1}(t)\right\rangle  & = & \frac{1}{\sqrt{2}}\left(e^{-i\left(\vartheta_{0}+\vartheta_{+}\right)}a_{+}^{\dagger}(0)+e^{i\left(\vartheta_{0}+\vartheta_{-}\right)}a_{-}^{\dagger}(0)\right)\nonumber \\
 &  & \times c_{+}^{\dagger}v_{2-}^{\dagger}\left|0\right\rangle \\
 & = & \frac{e^{-i\left(\frac{\vartheta_{+}-\vartheta_{-}}{2}\right)}}{\sqrt{2}}\left(e^{-i\left(\vartheta_{0}+\frac{\vartheta_{+}+\vartheta_{-}}{2}\right)}a_{+}^{\dagger}(0) \right.
\nonumber\\
& & +\left.e^{i\left(\vartheta_{0}+\frac{\vartheta_{+}+\vartheta_{-}}{2}\right)}a_{-}^{\dagger}(0)\right) 
c_{+}^{\dagger}v_{2-}^{\dagger}\left|0\right\rangle 
\end{eqnarray}
 with $\vartheta_{+}(t)=-\frac{2g_{2}^{2}}{\delta}t$ and $\vartheta_{-}(t)=\left(\frac{2g_{1}^{2}}{\delta+\frac{\Delta_{S}}{\hbar}}-\frac{2g_{2}^{2}}{\delta}\right)t$,
resulting in a Faraday rotation angle of $\vartheta_{-1}(t)=\left[\vartheta_{+}(t)+\vartheta_{-}(t)\right]/2=-\left(\frac{2g_{2}^{2}}{\delta}-\frac{g_{1}^{2}}{\delta+\frac{\Delta_{S}}{\hbar}}\right)t$.
This result is in complete agreement with the result using Fermi's
golden rule above. If the initial e-h pair is $+1$ polarized, then
the state after time $t$ is given by 
\begin{eqnarray}
\left|\psi_{+1}(t)\right\rangle  & = & \frac{1}{\sqrt{2}}\left(e^{-i\left(\vartheta_{0}+\vartheta_{+}(t)\right)}a_{+}^{\dagger}(0)+e^{i\left(\vartheta_{0}+\vartheta_{-}(t)\right)}a_{-}^{\dagger}(0)\right)\nonumber \\
 &  & \times c_{-}^{\dagger}v_{2+}^{\dagger}\left|0\right\rangle \\
 & = & \frac{e^{-i\left(\frac{\vartheta_{+}-\vartheta_{-}}{2}\right)}}{\sqrt{2}}\left(e^{-i\left(\vartheta_{0}+\frac{\vartheta_{+}+\vartheta_{-}}{2}\right)}a_{+}^{\dagger}(0) \right.
\nonumber\\
& & +\left.e^{i\left(\vartheta_{0}+\frac{\vartheta_{+}+\vartheta_{-}}{2}\right)}a_{-}^{\dagger}(0)\right)
c_{-}^{\dagger}v_{2+}^{\dagger}\left|0\right\rangle 
\end{eqnarray}
 with $\vartheta_{+}(t)=-\left(\frac{2g_{1}^{2}}{\delta+\frac{\Delta_{S}}{\hbar}}-\frac{2g_{2}^{2}}{\delta}\right)t$
and $\vartheta_{-}(t)=\frac{2g_{2}^{2}}{\delta}t$, resulting in a
Faraday rotation angle of $\vartheta_{+1}(t)=\left[\vartheta_{+}(t)+\vartheta_{-}(t)\right]/2=+\left(\frac{2g_{2}^{2}}{\delta}-\frac{g_{1}^{2}}{\delta+\frac{\Delta_{S}}{\hbar}}\right)t$.
This result is in complete agreement with the result using Fermi's
golden rule above.

In addition, the quantum-optical calculation lets us entangle the
photon with the electron-hole state on the 3D TI QD. In particular,
if we choose the initial state to be 
\begin{equation}
\left|\psi(0)\right\rangle =\frac{1}{\sqrt{2}}\left(a_{+}^{\dagger}+a_{-}^{\dagger}\right)\left(c_{+}^{\dagger}v_{2-}^{\dagger}+c_{-}^{\dagger}v_{2+}^{\dagger}\right)\left|0\right\rangle 
\end{equation}
 the photon and the e-h pair get fully entangled for $\vartheta_{\mp1}(\tau)=\pm\frac{\pi}{4}$,
i.e. after a time $\tau=\pi/4\left(\frac{2g_{2}^{2}}{\delta}-\frac{g_{1}^{2}}{\delta+\frac{\Delta_{S}}{\hbar}}\right)$,
yielding 
\begin{eqnarray}
\left|\psi(\tau)\right\rangle  & = & \frac{1}{\sqrt{2}}\left(e^{-i\frac{\pi}{4}}a_{+}^{\dagger}(0)+e^{i\frac{\pi}{4}}a_{-}^{\dagger}(0)\right)c_{+}^{\dagger}v_{2-}^{\dagger}\\
 &  & +\frac{1}{\sqrt{2}}\left(e^{-i\left(-\frac{\pi}{4}\right)}a_{+}^{\dagger}(0)+e^{i\left(-\frac{\pi}{4}\right)}a_{-}^{\dagger}(0)\right)c_{-}^{\dagger}v_{2+}^{\dagger}\left|0\right\rangle \nonumber 
\end{eqnarray}
 This state consists of a photon entangled to the e-h pair on the
3D TI QD.

\begin{figure}
\includegraphics[width=8.5cm]{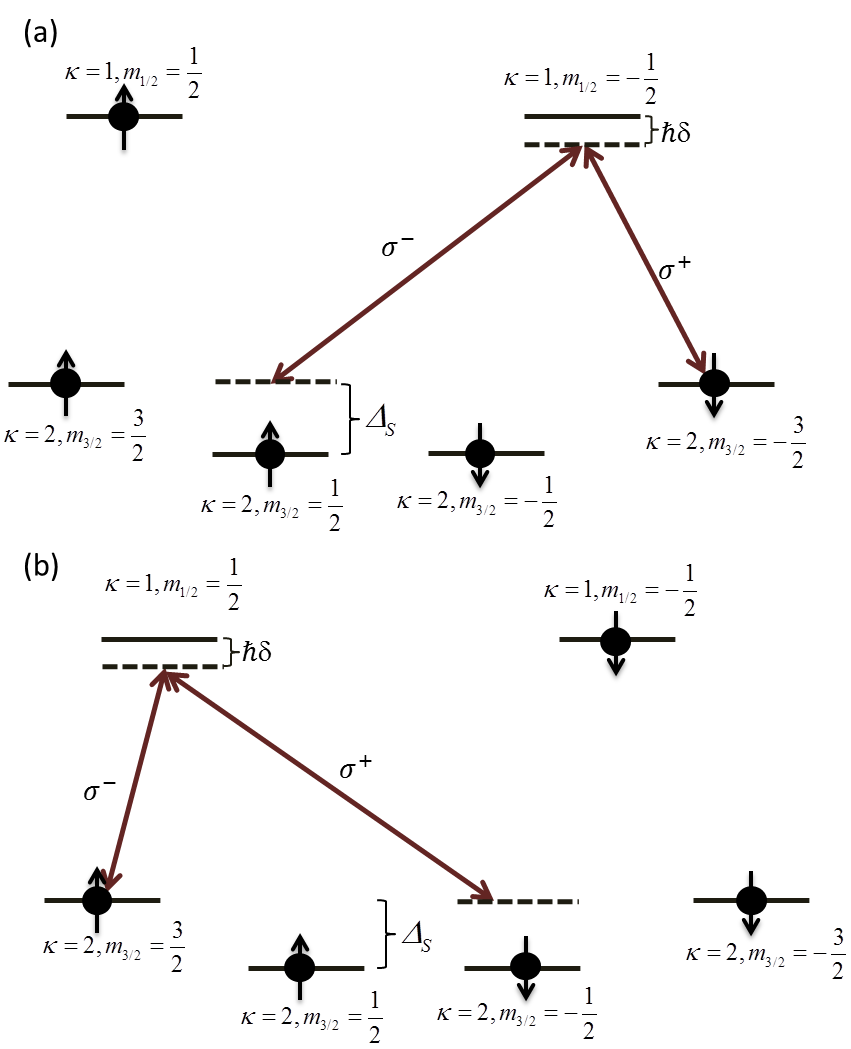}\caption{This is one possible level configuration that can be used for the
implementation of the quantum Faraday rotation, where the quantum
information is stored in form of an electron (a) in the spin up state
or (b) in the spin down state. (a) A single spin up electron is probed
by using off-resonant linearly polarized photon that acquires a positive
Faraday rotation angle through virtual excitation of e-h pairs. (b)
A single spin down electron is probed by using off-resonant linearly
polarized photon that acquires a negative Faraday rotation angle through
virtual excitation of e-h pairs.}

\label{fig:FaradayRotation_hole} 
\end{figure}

\begin{figure}
\includegraphics[width=8.5cm]{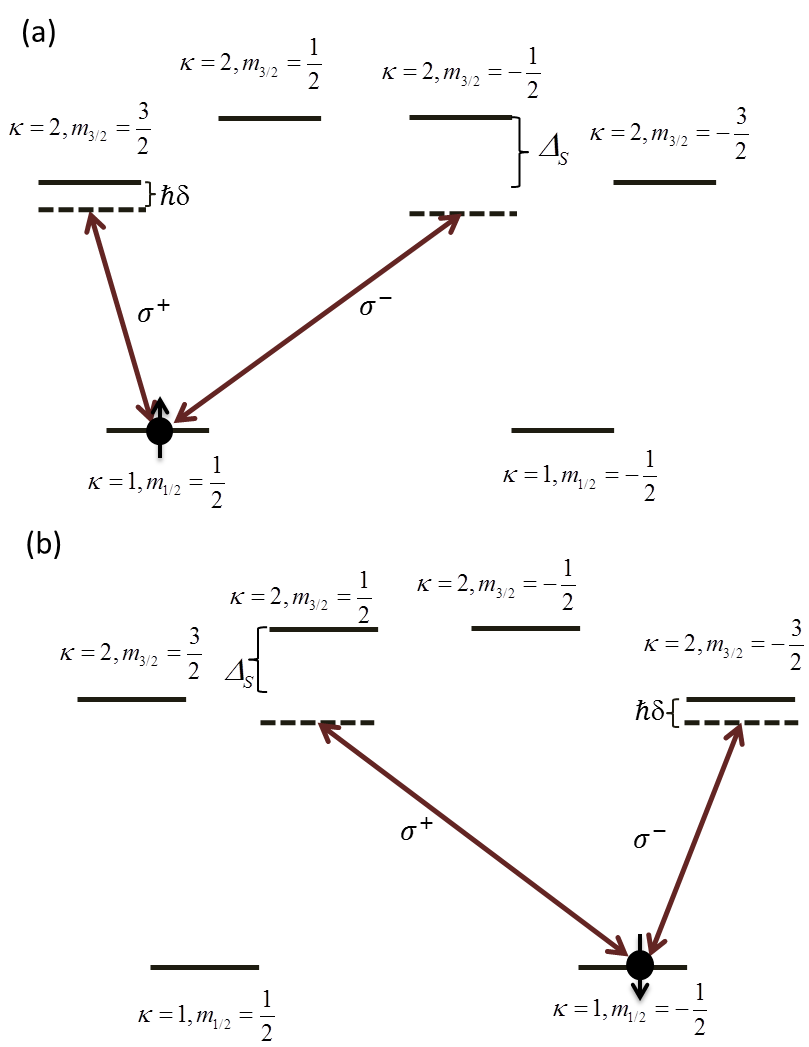}\caption{This is another possible level configuration that can be used for
the implementation of the quantum Faraday rotation, where the quantum
information is stored in form of a hole (a) in the spin up state or
(b) in the spin down state. (a) A single spin up hole is probed by
using off-resonant linearly polarized photon that acquires a positive
Faraday rotation angle through virtual excitation of e-h pairs. (b)
A single spin down hole is probed by using off-resonant linearly polarized
photon that acquires a negative Faraday rotation angle through virtual
excitation of e-h pairs.}

\label{fig:FaradayRotation_electron} 
\end{figure}

We consider two possible level configurations due to the electron-hole
symmetry in 3D TI QDs: 
\begin{enumerate}
\item Fig. \ref{fig:FaradayRotation_hole} shows the first level configuration,
in which only one of the states $\left|\Phi_{\frac{1}{2},\pm\frac{1}{2}}^{\kappa=1}\right\rangle $
is populated with an electron. 
\item Fig. \ref{fig:FaradayRotation_electron} shows the second level configuration,
in which only one of the states $\left|\Phi_{\frac{1}{2},\pm\frac{1}{2}}^{\kappa=1}\right\rangle $
is populated with a hole. 
\end{enumerate}

\subsection{Photon polarization - electron spin entanglement}

Now let us consider the Faraday effect due to a single electron for
the level configuration shown in Fig. \ref{fig:FaradayRotation_electron}.
Here the electron is in a s-like state. If the initial state is 
\begin{equation}
\left|\psi(0)\right\rangle =\frac{1}{\sqrt{2}}\left(e^{-i\vartheta_{0}}a_{+}^{\dagger}+e^{i\vartheta_{0}}a_{-}^{\dagger}\right)c_{+}^{\dagger}\left|0\right\rangle 
\end{equation}
 then state after the interaction is given by 
\begin{eqnarray}
\left|\psi(t)\right\rangle & = & \frac{1}{\sqrt{2}}\left(e^{-i\left(\vartheta_{0}-\frac{2g_{2}^{2}}{\delta}t\right)}a_{+}^{\dagger}(0) \right.
\nonumber\\
& & +\left.e^{i\left(\vartheta_{0}+\frac{2g_{1}^{2}}{\delta+\Delta_{S}}t\right)}a_{-}^{\dagger}(0)\right) c_{+}^{\dagger}\left|0\right\rangle 
\end{eqnarray}
 resulting in $\vartheta_{+1}(t)=\left(\frac{g_{1}^{2}}{\delta+\frac{\Delta_{S}}{\hbar}}-\frac{g_{2}^{2}}{\delta}\right)t$.
Conversely, if the initial state is (see Fig.\ref{fig:FaradayRotation_electron}
(b)) 
\begin{equation}
\left|\psi(0)\right\rangle =\frac{1}{\sqrt{2}}\left(e^{-i\vartheta_{0}}a_{+}^{\dagger}+e^{i\vartheta_{0}}a_{-}^{\dagger}\right)c_{-}^{\dagger}\left|0\right\rangle 
\end{equation}
 then state after the interaction is given by 
\begin{eqnarray}
\left|\psi(t)\right\rangle & = & \frac{1}{\sqrt{2}}\left(e^{-i\left(\vartheta_{0}-\frac{2g_{1}^{2}}{\delta+\frac{\Delta_{S}}{\hbar}}t\right)}a_{+}^{\dagger}(0) \right.
\nonumber\\
& & +\left.e^{i\left(\vartheta_{0}+\frac{2g_{2}^{2}}{\delta}t\right)}a_{-}^{\dagger}(0)\right) c_{-}^{\dagger}\left|0\right\rangle 
\end{eqnarray}
 resulting in $\vartheta_{-1}(t)=-\left(\frac{g_{1}^{2}}{\delta+\frac{\Delta_{S}}{\hbar}}-\frac{g_{2}^{2}}{\delta}\right)t$.
This result is also in complete agreement with the Faraday effect
obtained above.

Again, the quantum-optical calculation allows us to entangle the photon
with the single electron on the 3D TI QD. In particular, if we choose
the initial state 
\begin{equation}
\left|\psi(0)\right\rangle =\frac{1}{\sqrt{2}}\left(a_{+}^{\dagger}+a_{-}^{\dagger}\right)\left(c_{+}^{\dagger}+c_{-}^{\dagger}\right)\left|0\right\rangle 
\end{equation}
 then the state after the interaction is fully entangled after a time
$\tau=\pi/\left[4\left(\frac{g_{1}^{2}}{\delta+\frac{\Delta_{S}}{\hbar}}-\frac{g_{2}^{2}}{\delta}\right)\right]$,
i.e. 
\begin{eqnarray}
\left|\psi(\tau)\right\rangle  & = & \frac{1}{\sqrt{2}}\left(e^{-i\frac{\pi}{4}}a_{+}^{\dagger}(0)+e^{i\frac{\pi}{4}}a_{-}^{\dagger}(0)\right)c_{+}^{\dagger}\\
 &  & +\frac{1}{\sqrt{2}}\left(e^{-i\left(-\frac{\pi}{4}\right)}a_{+}^{\dagger}(0)+e^{i\left(-\frac{\pi}{4}\right)}a_{-}^{\dagger}(0)\right)c_{-}^{\dagger}\left|0\right\rangle \nonumber 
\end{eqnarray}
 This state is a fully entangled electron-photon state.

The Faraday effect due to a single hole for the level configuration
shown in Fig. \ref{fig:FaradayRotation_hole} can be calculated in
a similar way. Here the hole is in an s-like state. 
Other possible configurations include a single electron in a p-like state
or a single hole in a p-like state.

Due to the symmetry between positive- and negative-energy solutions,
in a 3D TI QD it is possible to define an electron spin qubit in terms
of an s-like or a p-like state. At the same time it is possible to
define a hole spin qubit in terms of an s-like or a p-like state.
This cannot be done in a conventional wide-bandgap semiconductor QD,
where the electron is associated with an s-like state and the hole
is associated with a p-like state.\cite{Singh}

For a quantitative description, we can assume a single 3D TI QD embedded
in a semiconductor microcavity. The strong and weak interaction can
occur between the QD e-h pair and discretized cavity modes at resonance,
$\omega_{21}=\omega_{c}$. The e-h-photon coupling parameter g is
given by $g=\left(\pi e^{2}f\right)^{1/2}/\left(4\pi\epsilon_{r}\epsilon_{o}m_{o}V_{m}\right)^{1/2}$,
where $\epsilon_{r}$ is the dielectric constants for the cavity material,
\cite{Forchel} $m_{o}$ is the free electron mass, and $V_{m}$ is
the mode volume. The mode volume for a mode of wavelength $\lambda$
is $V_{m}\approx\left(\lambda/2n\right)^{3}$, where $n=\sqrt{\epsilon_{r}}$
(for a GaAs microcavity, $n=3.31$). Using $\hbar\omega_{21}=130$
meV, the oscillator strengths for the transition $\Phi_{\frac{3}{2},+\frac{1}{2}}^{\kappa=2}\longleftrightarrow\Phi_{\frac{1}{2},-\frac{1}{2}}^{\kappa=1}$
and $\Phi_{\frac{3}{2},-\frac{3}{2}}^{\kappa=2}\longleftrightarrow\Phi_{\frac{1}{2},-\frac{1}{2}}^{\kappa=1}$
are obtained, respectively, $f_{1}\thickapprox9$ and $f_{2}\thickapprox27$.
This gives us an estimate of $\hbar g_{1}\approx10\:\mu$eV and
$\hbar g_{2}\approx17\:\mu$eV. For a detuning energy
of $\hbar\delta=100\:\mu$eV the time it takes to fully entangle the
electron spin and the photon polarization is calculated to be of the
order of 180 ps. The necessary condition to be in the strong coupling regime is that
$g$ must be large compared to both spontaneous emission rate and
cavity decay loss rate.\cite{Fox} Thus, for $Q\geq\omega/g_{1}\thickapprox1.3\times10^{4}$
the 3D TI QD is in the strong coupling regime. For $Q=10^{5}$, the
photon decay rate is given by $\kappa=\frac{\omega}{2\pi Q}=3.1\times10^{9}$
s$^{-1}$. This gives a cavity photon life time of 3 ns.

\section{\label{Conclusions}Conclusions}

We have shown that Weyl fermions can be confined in all three dimensions
at the spherically shaped interface between two narrow-bandgap semiconductor
alloys, such as the core-bulk heterostructure made of PbTe/Pb$_{0.31}$Sn$_{0.69}$Te.
This configuration provides us with the model of a spherical 3D TI
QD with tunable size $r_{0}$ and potential $\Delta_{0}$, which allows
for complete control over the number of bound interface states. The
most important features of 3D TI have been identified in a 3D TI QD,
namely the spin locking effect and the Kramers degeneracy. We found
that the Weyl states are confined on the surface of the QD, in contrast
to the electrons and holes in topologically trivial semiconductor
QDs. We showed that due to the large dipole moment of 450 Debye it
is possible to reach the strong-coupling regime inside a cavity with
a quality factor of $Q\approx10^{4}$ in the infrared wavelength regime
around $10\:\mu$m. Because of the strict optical selection rules,
the 3D TI QD gives rise to interesting applications based on the semi-classical
and quantum Faraday effect. We found that the 3D TI QD is a good candidate
for quantum memory, quantum teleportation, and quantum computing
with single spins in 3D TI QDs using infrared light. In particular,
a single e-h pair, a single electron, or a single hole can be used
as a qubit for the implementation of optically mediated quantum computing
with 3D TI QDs. Interestingly, we found that due to the symmetry between
positive- and negative-energy solutions, in a 3D TI QD it is possible
to define an electron spin qubit in terms of an s-like or a p-like
state. At the same time it is possible to define a hole spin qubit
in terms of an s-like or a p-like state. This cannot be done in a
zincblende wide direct-bandgap semiconductor QD, where the electron
is associated with an s-like state and the hole is associated with
a p-like state.\cite{Singh}
\begin{acknowledgments}
We acknowledge support from NSF (Grants ECCS-0901784 and Grant ECCS-1128597)
and AFOSR (Grant FA9550-09-1-0450). M.N.L. thanks Daniel Loss for
fruitful discussions during his stay at the University of Basel, Switzerland.
M.N.L. acknowledges partial support from the Swiss National Science
Foundation. We thank Mikhail Erementchouk for useful discussions. 
\end{acknowledgments}
\appendix

\section{\label{sec:Wronskian}Calculation of the Wronskian }

The Wronskian of the functions $\mathcal{I_{\kappa}}\left(z\right)$
and $\mathcal{K_{\kappa}}\left(z\right)$ is defined as \cite{Abramowitz}
\begin{equation}
W_{\kappa}\left[\mathcal{I}_{\kappa}\left(z\right),\mathcal{K}_{\kappa}\left(z\right)\right]=\mathcal{I}_{\kappa}\left(z\right)\mathcal{K}_{\kappa}^{'}\left(z\right)-\mathcal{I}_{\kappa}^{'}\left(z\right)\mathcal{K}_{\kappa}\left(z\right),
\end{equation}
 where the prime denotes the derivative of the function. For independent
solutions, it is to be noted that Wronskian is proportional to $1/p\left(x\right)$
in a Sturm-Liouville type equation $\frac{d}{dx}\left[p\left(x\right)\frac{dy}{dx}\right]+g\left(x\right)y=0$.
Therefore, Wronskians in the text are calculated to be 
\begin{eqnarray}
W_{\kappa}\left[\mathcal{I}_{\kappa}\left(z\right),\mathcal{K}_{\kappa}\left(z\right)\right] & = & -\frac{1}{z^{2}},\\
W_{\kappa-1}\left[\mathcal{I}_{\kappa-1}\left(z\right),\mathcal{K}_{\kappa-1}\left(z\right)\right] & = & -\frac{1}{z^{2}}.
\end{eqnarray}

\section{Limiting form of Bessel functions }

\label{sec:Bessel}

The limiting forms of modified Bessel functions for $z\rightarrow0$
are given by 
\begin{equation}
\left.\begin{array}{c}
I_{\kappa}\left(z\right)=\frac{1}{\Gamma\left(\kappa+1\right)}\left(\frac{z}{2}\right)^{\kappa}\\
K_{\kappa}\left(z\right)=\frac{\Gamma\left(\kappa\right)}{2}\left(\frac{2}{z}\right)^{\kappa}
\end{array}\right\} \textrm{ as }z\rightarrow0.
\end{equation}
 The modified spherical Bessel functions can be written in terms of
modified Bessel functions as 
\begin{equation}
\mathcal{I_{\kappa}}\left(z\right)=\left(\sqrt{\frac{\pi}{2z}}\right)I_{\kappa+\frac{1}{2}}\left(z\right),\:\mathcal{K_{\kappa}}\left(z\right)=\left(\sqrt{\frac{2}{\pi z}}\right)K_{\kappa+\frac{1}{2}}\left(z\right).
\end{equation}
 Therefore, the function 
\begin{equation}
F\left(z\right)=\left[z\mathcal{I}_{\kappa}\left(z\right)\mathcal{K}_{\kappa}\left(z\right)\right]\left[z\mathcal{I}_{\kappa-1}\left(z\right)\mathcal{K}_{\kappa-1}\left(z\right)\right].
\end{equation}
 has the limiting form $F\left(z\right)=\frac{1}{4\kappa^{2}-1}$
as $z\rightarrow0$.

The asymptotic expansion $\left(z\rightarrow\infty\right)$ of the
modified Bessel functions are given by

\begin{equation}
\begin{array}{c}
I_{\kappa}\left(z\right)=\frac{e^{z}}{\sqrt{2\pi z}}\left\{ 1-\frac{4\kappa^{2}-1}{8z}+\frac{\left(4\kappa^{2}-1\right)\left(4\kappa^{2}-9\right)}{2!\left(8z^{2}\right)}-...\right\} \\
K_{\kappa}\left(z\right)=\sqrt{\frac{\pi}{2z}}e^{-z}\left\{ 1+\frac{4\kappa^{2}-1}{8z}+\frac{\left(4\kappa^{2}-1\right)\left(4\kappa^{2}-9\right)}{2!\left(8z^{2}\right)}-...\right\} 
\end{array}.
\end{equation}

\section{The fermion doubling theorem}

\label{sec:doubling}

Nielsen and Ninomiya investigated Weyl fermions on a crystal.\cite{Nielsen&Ninomiya}
They formulated a no-go theorem, called the fermion doubling theorem,
requiring that Weyl nodes in a crystal always exist in pairs of opposite
chirality. The reason for this theorem is that the number of Weyl
fermions in the first Brillouin zone must be conserved. This conservation
law can be checked by calculating the Berry flux in the first Brillouin
zone. 

It is important to note that the fermion doubling theorem is only valid for continuum states. 
Therefore it does not apply to the bound eigenstates of the 3D TI QD, which have a discrete eigenspectrum.
Below we give arguments for the validity of the fermion doubling theorem in the continuum limit,
which corresponds to the asymptotic limit when the 3D TI QD radius $r_o$ becomes infinite.

A typical calculation of the Berry curvature $\mathcal{B}_{n}(\mathbf{k})=\nabla_{\mathbf{k}}\times\mathcal{A}_{n}(\mathbf{k})$
considers a single band Bloch state $u_{n\mathbf{k}}(\mathbf{r})e^{i\mathbf{k\cdot r}}$,
which gives rise to the Berry connection $\mathcal{A}_{n}(\mathbf{k})=i\int_{\Omega}d^{3}r\; u_{n\mathbf{k}}^{*}(\mathbf{r})\nabla_{\mathbf{k}}u_{n\mathbf{k}}(\mathbf{r})$.\cite{Marder}
As long as the nth band does not touch or cross any other band, the
Berry flux is zero, i.e. $\nabla_{\mathbf{k}}\cdot\mathcal{B}_{n}(\mathbf{k})=0$
. However, if there is a band crossing, this situation changes drastically
due to the monopole at the crossing point. Using $\mathbf{k}\cdot\mathbf{p}$
approximation, around the crossing point in the first Brillouin zone
the Berry connection becomes $\mathcal{A}_{\pm}(\mathbf{k})=i\left\langle \chi_{\pm}\right|\nabla_{\mathbf{k}}\left|\chi_{\pm}\right\rangle $,
where $\chi_{\pm}$ is the four-spinor of the solution $\Phi_{\pm}=\chi_{\pm}F(\mathbf{r})$
of Eq. (\ref{eq:1}).\cite{Paudel&Leuenberger} Assuming a very large
QD, where quantum confinement can be neglected, the four-spinor reads
\begin{equation}
\chi_{\pm}=\left(\begin{array}{c}
\pm e^{-i\frac{\left(\varphi\pm\pi/2\right)}{2}}\\
\pm e^{i\frac{\left(\varphi\pm\pi/2\right)}{2}}\\
e^{-i\frac{\left(\varphi\mp\pi/2\right)}{2}}\\
e^{i\frac{\left(\varphi\mp\pi/2\right)}{2}}
\end{array}\right)
\end{equation}
 where $e^{\mp i\varphi}=\frac{k_{x}\mp ik_{y}}{k_{\perp}}$ and the
position-dependent function is given by $F(\mathbf{r})=Ce^{-\frac{1}{\hbar v_{\parallel}}\overset{z}{\underset{0}{\int}}\Delta(z^{'})dz^{'}+i\mathbf{k}_{\perp}\cdot\mathbf{r}}$,
where $C$ is the normalization constant. In order to capture the
Berry curvature apart from the azimuthal angle $\varphi$ we need
to add the dependence on the polar angle $\theta$. At the same time,
we perform the gauge transformations $e^{\pm i\varphi}$ to shift
the singularity of the Berry curvature to the south pole. This means
we calculate the Berry curvature with respect to the normalized 4-spinors
\begin{eqnarray}
\chi_{C,+} & = & \frac{1}{\sqrt{2}}\left(\begin{array}{c}
e^{-i\frac{\pi}{4}}\cos\frac{\theta}{2}\\
e^{i\left(\varphi+\frac{\pi}{4}\right)}\sin\frac{\theta}{2}\\
e^{i\frac{\pi}{4}}\cos\frac{\theta}{2}\\
e^{i\left(\varphi-\frac{\pi}{4}\right)}\sin\frac{\theta}{2}
\end{array}\right),\nonumber \\
\chi_{C,-} & = & \frac{1}{\sqrt{2}}\left(\begin{array}{c}
-e^{-i\left(\varphi-\frac{\pi}{4}\right)}\sin\frac{\theta}{2}\\
e^{-i\frac{\pi}{4}}\cos\frac{\theta}{2}\\
e^{-i\left(\varphi+\frac{\pi}{4}\right)}\sin\frac{\theta}{2}\\
-e^{i\frac{\pi}{4}}\cos\frac{\theta}{2}
\end{array}\right)
\end{eqnarray}
 The Berry connection is then given by 
\begin{equation}
\mathcal{A}_{\pm}(\mathbf{k})=i\left\langle \chi_{C,\pm}\right|\nabla_{\mathbf{k}}\left|\chi_{C,\pm}\right\rangle =\mp\frac{(1-\cos\theta)}{2k\sin\theta}\mathbf{e}_{\varphi}
\end{equation}
 where $\mathbf{e}_{\varphi}$ is the unit vector pointing in $\varphi$-direction.
Thus, we obtain the Berry phase 
\begin{equation}
\gamma_{\pm}=\oint\mathcal{A}_{\pm}(\mathbf{k})\cdot d\mathbf{k}=\mp\pi(1-\cos\theta)
\end{equation}
 and the Berry curvature 
\begin{equation}
\mathcal{B}_{\pm}(\mathbf{k})=\mp\frac{1}{2k^{2}}\mathbf{e}_{k}\label{eq:Berry_curvature_k}
\end{equation}
 Note that the Berry curvature for the 4-spinor is the same as the
Berry curvature of a 2-spinor.\cite{Shankar} For a loop on the 2D
surface where $\theta=\pi/2$, we get $\gamma_{\pm}=\mp\pi$, which
gives rise to the topological phase shift seen in Shubnikov-de Haas
oscillations for the surface of 3D topological insulators.\cite{Veldhorst}
From $\triangle(1/k)=\mp4\pi\delta^{(3)}(\mathbf{k})$ and $\nabla(1/k)=\mp\frac{1}{k^{2}}\mathbf{e}_{k}$
it follows that the Berry curvature is the solution of the equation
\begin{equation}
\nabla_{\mathbf{k}}\cdot\mathcal{B}_{\pm}(\mathbf{k})=\mp4\pi g\delta^{(3)}(\mathbf{k})
\end{equation}
 where $g=\mp1/2$ is the strength of the Dirac monopole for positive
and negative helicity of the 4-spinor, which is identical to the result
for 2-spinors (see Refs.~\onlinecite{Shankar} and \onlinecite{Nakahara}).

In order to understand the helicity of the Weyl fermions at the interface,
we have shown in Ref.~\onlinecite{Paudel&Leuenberger} that the
helicity operator is given by 
\begin{equation}
\hat{h}_{\textrm{TI}}=\left(1/\left|p_{\bot}\right|\right)\left(\begin{array}{cc}
\left(\boldsymbol{\sigma}_{\perp}\times\boldsymbol{p}_{\perp}\right)\cdot\boldsymbol{\hat{z}} & 0\\
0 & -\left(\boldsymbol{\sigma}_{\perp}\times\boldsymbol{p}_{\perp}\right)\cdot\boldsymbol{\hat{z}}
\end{array}\right)
\end{equation}
 which commutes with the Hamiltonian in Eq.~(\ref{eq:1}) and yields
$\hat{h}_{\mathrm{TI}}\Phi_{\pm}=\left(\pm1/2\right)\Phi_{\pm}$,
where the + sign denotes the positive helicity of positive-energy
solutions and the - sign denotes the negative helicity of negative-energy
solutions. This provides the possibility to write an effective 2D
Hamiltonian for the Weyl fermions on the surface of 3D topological
insulators, i.e. 
\begin{equation}
H_{2D}=\hbar v\left(\begin{array}{cc}
\left(\boldsymbol{\sigma}_{\perp}\times\boldsymbol{k}_{\perp}\right)\cdot\boldsymbol{\hat{z}} & 0\\
0 & -\left(\boldsymbol{\sigma}_{\perp}\times\boldsymbol{k}_{\perp}\right)\cdot\boldsymbol{\hat{z}}
\end{array}\right)
\end{equation}
 This effective 2D Hamiltonian can be reduced to two Weyl Hamiltonians
of the form $H_{2D}^{2x2}=\pm\hbar v\left(\boldsymbol{\sigma}_{\perp}\times\boldsymbol{k}_{\perp}\right)\cdot\boldsymbol{\hat{z}}$.
It is important to note that both 2-spinors of $\chi_{\pm}$, the
2-spinor $\chi_{\pm}^{L^{-}}$ of the $L^{-}$ band and the 2-spinor
$\chi_{\pm}^{L^{+}}$ of the $L^{+}$ band have the same helicity,
in contrast to the commonly used Weyl Hamiltonians $H_{W}(\mathbf{k})=\pm\hbar v\mathbf{\mathbf{\mathbf{\boldsymbol{\sigma}\cdot}k}}$.
The reason for this is that the two 2-spinors are coupled through
the mass term $\Delta(z)$ in $z$-direction, as given in the 3D Hamiltonian
in Eq.~(\ref{eq:1}).

\begin{figure}
\includegraphics[width=8.5cm]{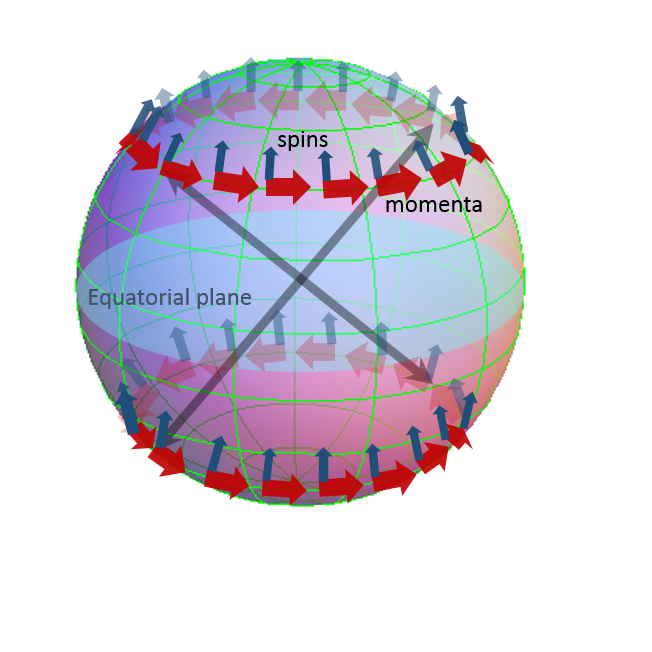}\caption{Two antipodal points on the surface of the sphere defined by the QD
are identified as the Dirac cones of opposite helicity. One point
lies on the northern hemisphere, while its antipodal point lies on
the southern hemisphere. The currents flowing along the latitudes can be imagined  
as angular momentum states of a 3D TI QD in the continuum limit. At
the antipodal points momenta (red arrows) point in opposite $\hat{\varphi}$
direction to each other while spins (blue arrows) point in the same
$\hat{\theta}$ direction, where $\hat{\theta}$ and $\hat{\varphi}$
are the spherical angular unit vectors. Hence, they have opposite
chirality. This satisfies the fermion doubling theorem.}

\label{fig:doubling_theorem} 
\end{figure}

In order to satisfy the fermion doubling theorem,\cite{Nielsen&Ninomiya}
usually the Dirac cones on the opposite side of the slab of a 3D topological
insulator are identified as the fermion doublers. In the case of the
3D IT QD, for $r_{o}\rightarrow\infty$, i.e. in the continuum limit, the Berry curvature in $\mathbf{k}$-space
for a 2D interface, given by Eq.~(\ref{eq:Berry_curvature_k}), determines
the Weyl nodes that need to satisfy the fermion doubling theorem.
Hence, according to Ref.~\onlinecite{Lee:2009}, we can adopt the mapping of the two opposite surfaces of a
3D slab of TI onto the northern and southern hemispheres of a sphere.
We then identify the pairs of Dirac cones with opposite helicity as
the ones located on the antipodal points on the surface of the sphere
defined by the QD, as shown in Fig.~\ref{fig:doubling_theorem}.
Note that in both cases, the slab and the QD, the pairs of Dirac cones
map into each other through the parity transformation, which in general
reverses the helicity. We can identify a current
on the surface of the sphere flowing along a latitude. The parity
transformation then maps one latitude on the northern hemisphere with
one type of helicity to its partner latitude on the southern hemisphere
with the opposite helicity. These arguments show that the fermion
doubling theorem is satisfied for a 3D TI QD in the continuum limit.

\end{document}